\pdfoutput=1
\documentclass[%
 aip,
 amsmath,amssymb,
 reprint,%
]{revtex4-1}

\usepackage{graphicx}
\usepackage{dcolumn}
\usepackage{bm}
\usepackage{xcolor}
\usepackage[utf8]{inputenc}
\usepackage[T1]{fontenc}
\usepackage{mathptmx}
\usepackage{etoolbox}
\usepackage{makecell}
\makeatletter
\def\@email#1#2{%
 \endgroup
 \patchcmd{\titleblock@produce}
  {\frontmatter@RRAPformat}
  {\frontmatter@RRAPformat{\produce@RRAP{*#1\href{mailto:#2}{#2}}}\frontmatter@RRAPformat}
  {}{}
}%
\makeatother
\begin{document}

\preprint{AIP/123-QED}

\title[Bianco et al. ]{Heart Rate Variability Series is the Output of a non-Chaotic System driven by Dynamical Noise}
\author{M. Bianco}
\author{A. Scarciglia}%
\affiliation{
Dept. of Information Engineering and Research Centre E Piaggio, University of Pisa, Pisa, Italy
}%
\author{C. Bonanno}
\affiliation{%
Dept. of Mathematics, University of Pisa, Pisa, Italy 
}%
\author{G. Valenza}
\affiliation{
Dept. of Information Engineering and Research Centre E Piaggio, University of Pisa, Pisa, Italy
}%

\date{\today}

\begin{abstract}
Heart rate variability (HRV) series reflects the dynamical variation of heartbeat-to-heartbeat intervals in time and is one of the outputs of the cardiovascular system. Over the years, this system has been recognized for generating nonlinear and complex heartbeat dynamics, with the latter referring to a high sensitivity to small —theoretically infinitesimal— input changes. While early research associated chaotic behavior with the cardiovascular system, evidence of stochastic inputs to the system, i.e., a physiological noise, invalidated those conclusions. To date, a comprehensive characterization of the cardiovascular system dynamics, accounting for dynamical noise input, has not been undertaken. In this study, we propose a novel methodological framework for evaluating the presence of regular or chaotic dynamics in noisy dynamical systems. The method relies on the estimation of asymptotic growth rate of noisy mean square displacement series in a two-dimensional phase space. We validated the proposed method using synthetic series comprising well-known regular and chaotic maps. We applied the method to real HRV series from healthy subjects, as well as patients with atrial fibrillation and congestive heart failure, during unstructured long-term activity. Results indicate that HRV series are consistently generated by a regular system driven by dynamical noise.
\end{abstract}

\maketitle

\begin{quotation}
We address previous misconceptions surrounding chaotic behavior in the cardiovascular system, emphasizing the role of stochastic dynamical inputs. The proposed methodological framework provide a novel mean to characterizing physiological dynamics in terms of regular versus chaotic patterns. While cardiac pathology does not modulate chaotic behavior, atrial fibrillation induces higher sensitivity to input changes. This research contributes  to our understanding of cardiovascular dynamics. 
\end{quotation}

\section{\label{sec:level1}Introduction}
The dynamics of time intervals between successive heartbeats, known as the heart period, define the series known as Heart Period Variability series. This heartbeat dynamics consists of successive RR intervals derived from the electrocardiogram (ECG) and represents an output of the cardiovascular system. While heart period and heart rate are inversely related, Heart Period Variability series are commonly referred to as Heart Rate Variability (HRV) series. Current understanding acknowledges that HRV originates from a multitude of inputs converging at the sinoatrial node. Predominantly, the sympathetic and parasympathetic nervous systems play pivotal roles, alongside respiration, blood pressure, and hormonal influences, collectively modulating cardiac autonomic activity \cite{akselrod1981power,saul2021heart,rajendra2006heart,stauss2003heart,billman2011heart,cygankiewicz2013heart,van1993heart,malik1998heart}.

Pioneering studies analyzed short-term HRV series in the frequency domain, spanning 3–5 minutes recordings \cite{akselrod1981power,rajendra2006heart,de1985relationships,saul1989transfer,saul1991cardiorespiratory,saul1988nonlinear}. These investigations unveiled three characteristic peaks in the HRV power spectrum: a high-frequency (HF) peak often aligning with the respiratory frequency, leading to a former definition of respiratory sinus arrhythmia; a low frequency (LF) peak around 0.1 Hz, whose amplitude can be modulated by arterial blood pressure and sympathovagal activity; and, often, a peak around 0.04Hz that has formerly been linked to peripheral vasomotor regulation \cite{saul2021heart, porta2002quantifying, madwed1989low, madwed1991heart, baselli1994model, porta2000assessing}. Accordingly, HRV spectrum has been divided into three frequency bands: a HF band (0.15 to 0.4 Hz), predominantly reflective of respiratory modulation of cardiac vagal activity; a LF band (0.04 to 0.15 Hz), reflective of both sympathetic and parasympathetic activities, as well as baroreflex feedback \cite{de1985relationships,saul1989transfer,valenza2018measures}; and a very low-frequency (VLF) band, below approximately 0.04 Hz, whose modulating physiological correlates are yet unknown.

Beyond the frequency domain, several studies have demonstrated the importance of accounting for heartbeat nonlinear dynamics for a comprehensive characterization of the cardiovascular system \cite{saul1991cardiorespiratory,saul1988nonlinear,sunagawa1998dynamic,glass2009introduction,spasic2019nonlinearity,barbieri2017complexity,sassi2015advances}.
A common methodological strategy involves constructing an M-dimensional trajectory vector \(X_M(t)\) by taking delayed samples of the HRV time series \(x(t)\), such that \(X_M(t)=[x(t), x(t + L), ..., x(t + ML - L)]\), where \(L\) is a selected fixed lag \cite{takens2006detecting}. Subsequently, indicators such as the largest Lyapunov exponent and the correlation dimension \(D_2\) may be estimated. The largest Lyapunov exponent represents the average exponential growth rate of the initial distance between two neighboring points in \(X_M(t)\) as time evolves \cite{rosenstein1993practical}, while the correlation dimension \(D_2\) measures the fractal dimension of the manifold enclosing the trajectories of the nonlinear dynamical system \cite{grassberger1983measuring,kantz2004nonlinear}. Other important estimators include the approximate entropy (ApEn) and the sample entropy (SampEn). ApEn \cite{pincus1991approximate,pincus1994physiological} was initially devised as a practical implementation of the Kolmogorov–Sinai entropy of a nonlinear dynamical system, while SampEn \cite{richman2000physiological} was introduced aiming to overcome ApEn numerical issues. Both metrics assess the likelihood that runs of templates that are close for \(m\) points remain close (less than a certain tolerance level \(r\)) for \(m + 1\) points in a given sequence of length \(N\). 

In general, many computational methods, including but not limited to the aforementioned ones, aim to assess and quantify the degree of complexity in unknown dynamical systems \cite{sassi2015advances}. While numerical issues may arise from the parameters selection and insufficient series cardinality, many complexity measurements are sensitive to the degree of randomness in the system under study \cite{li2015assessing,scarciglia2023estimation,scarciglia2024physiological}. This is especially evident for ApEn and SampEn, which then are deemed to quantify the irregularity of a series \emph{per se}. Hence, the increase in entropy may not be necessarily indicative of an increase in complexity \cite{li2015assessing,scarciglia2023estimation,scarciglia2024physiological}. In this sense, \emph{complexity} refers to the presence of nonrandom fluctuations, maybe on multiple time scales, in a seemingly irregular dynamics \cite{manor2010physiological}. Another good definition of system complexity involves the system composition; in this regard, a complex dynamical system is a nonlinear system as a whole exhibits properties across different domains that individual sub-components acting alone cannot fully capture, opening the definition to network dynamics \cite{san2023frontiers}.

Some complex systems may also exhibit chaotic dynamics. The definition of chaos requires deterministic dynamics generating trajectories in phase space that diverge exponentially over time. Therefore, chaos refers to aperiodic dynamics in deterministic systems with bounded dynamics and sensitive dependence on initial conditions \cite{glass2009introduction}. This sensitivity to initial conditions leads to trajectories diverging over time, often exhibiting wandering over a fractal geometric entity \cite{sassi2015advances}.

While nonlinearity has been experimentally demonstrated in the cardiovascular system \cite{sunagawa1998dynamic}, with particular emphasis on its nonlinear autonomic control, greater challenges arise when attempting to assess the presence of chaotic dynamics through the analysis of HRV series. In 2009, an informative essay introduced the question "\emph{Is the normal Heart Rate Chaotic?}" as a controversial topic in nonlinear science \cite{glass2009introduction}. Most studies cited therein were inconclusive, partly due to the mechanisms underlying HRV generation that include stochastic processes at the cellular level, the influence of respiration on heart rate, and the interactions of the multiple feedback loops regulating the cardiovascular system \cite{sunagawa1998dynamic}. 
To this end, a methodological framework capable of discerning regular versus chaotic dynamics in noisy dynamical systems is greatly needed. Indeed, HRV series are driven by a significant presence of dynamical physiological noise \cite{scarciglia2024physiological}. To illustrate, mathematical models elucidating the generation of normal sinus rhythm have revealed that the stochastic release of the regulatory agent acetylcholine in the vicinity of the SA node may induce an irregular rhythm, potentially misidentified as chaotic \cite{zhang2009stochastic}. 

In order to provide a comprehensive characterization of the cardiovascular system in terms of chaotic dynamics, we took inspiration from previous endeavors on chaos assessment \cite{gottwald20160,gottwald2009implementation} and here propose a methodology to quantitatively evaluate the chaotic nature of HRV series as driven by dynamical noise.

\section{Proposed Methodological Framework}
The proposed methodological framework allows for a quantitative assessment of chaotic and regular dynamics in noisy dynamical systems utilizing time series data. 

Let us consider a physiological system as a discrete dynamical system $(X, \nu, T )$, where $X$ denotes a compact subset of $\mathbb{R}$, and $\nu$ represents an ergodic probability measure preserved by the map $T$. A typical output of this system is denoted as 
\begin{equation}
\label{dyn_noise}
    x(n)=T(x(n-1), x(n-2), ..., x(0)) + \epsilon(n)
\end{equation} where $x(i)\in X$ for all positive integers $i$, and $\{\epsilon(n)\}_n$ is the dynamical physiological noise modeled as a sequence of IID Gaussian random variables with $\mathcal{N}(0,\sigma^2)$.
In this frame, assuming the map $T$ is deterministic and differentiable, the system dynamics can be either regular or chaotic. Given a  time series $\{x(n)\}$, perturbed by dynamical noise $\epsilon$ according to eq. (\ref{dyn_noise}) with $n=1,...,N$, we aim to calculate a value $K$ that is $0$ in the case of regular dynamics and $1$ in the case of chaotic dynamics. To this end, we define a 2-dimensional phase space as follows:

\[
\begin{cases}
p(n+1)=p(n)+x(n)\cos(cn)\\
q(n+1)=q(n)+x(n)\sin(cn)
\end{cases}
\]
where the normalized frequency $c\in (0,2\pi)$ is fixed.
Accordingly, in the regular case, the trajectories of $p(n)$ and $q(n)$ are typically bounded, whereas in the chaotic case, they typically behave approximately like a two-dimensional Brownian motion, evolving with a growth rate $\sqrt{n}$ (diffusively) \cite{gottwald20160}. To distinguish such behaviors, we introduce the \textit{time-averaged mean square displacement}:

\[
M(n)=\lim_{N\to \infty}\frac{1}{N}\sum_{j=1}^N ([p( j + n) - p( j)]^2 + [q( j + n) - q( j)]^2)
\]

$M(n)$ is bounded if the trajectories of $p(n)$ and $q(n)$ are bounded and grows linearly if they evolve diffusively \cite{gottwald20160}. 
Finally, we define:

\[
K =\lim_{n \to \infty}\frac{\log M(n)}{\log n}
\]

which captures the growth rate of $M(n)$. Hence, if $K = 0$ (for almost every choice of $c$), the dynamics are regular, while $K = 1$ (for almost every choice of $c$) signifies chaotic dynamics.

\subsection{Estimation algorithm}
Given an HRV series $\{RR(j)\}$ with $j=1,...,N$:

\begin{enumerate}
    \item For a fixed $c \in (0,2\pi)$, compute the quantities
    \[
    p_c(n)=\sum_{j=1}^n RR(j) \cos(jc) \quad \text{and} \quad q_c(n)=\sum_{j=1}^n RR(j) \sin(jc)
    \]
    for $n=1,...,N$.
    
    \item Then compute the mean square displacement
    \[
    M_c(n)=\frac{1}{N}\sum_{j=1}^N [p_c(j+n)-p_c(j)]^2-[q_c(j+n)-q_c(j)]^2
    \]
    for $n=1,...,n_{\text{cut}}$.
    \item Compute the modified mean square displacement by removing an oscillatory component
    \[
    D_c(n)=M_c(n)- \frac{1}{N}\sum_{j=1}^N RR(j) \frac{1-\cos(nc)}{1-\cos c} 
    \]
    for $n=1,...,n_{\text{cut}}$. 
    
     \item Then, add an oscillatory component whose amplitude is modulated by an Inverse Gaussian probability density function \cite{valenza2013point} of parameter ($\ell,m$) driven by the system's dynamical noise:  
    \[
    \Tilde{D}_c (n)=D_c(n)+ \sqrt{\frac{\ell}{2\pi \beta^3}} \exp\left(-\frac{\ell(\beta-m)^2}{2m^2\beta}\right) \sin(\sqrt 2)
    \]
    where $\beta=n_{\text{cut}}\frac{[RR]}{\sigma}$, where $[RR]$ is the series range.
       
    \item Compute the asymptotic growth rate $K_c$ of $\Tilde{D}_c(n)$ as the linear correlation coefficient between the vectors $u=[1,2,...,n_{\text{cut}}]$ and  $\Tilde{\Delta}=[\Tilde{D}_c(1),...,\Tilde{D}_c(n_{\text{cut}})]$:
    \[
    K_c=\frac{\text{Cov}(u,\Tilde{\Delta})}{\sqrt{\text{Var}(u)\text{Var}(\Tilde{\Delta})}}
    \]
    
    \item Perform the previous steps in an interval centred in $c$ and compute:
    \[
    K=\text{median}(K_c)
    \]
    considering all $K$ estimates.
\end{enumerate}

In this study, we set $\ell=2$ and $m=3$ from early evidence on synthetic datasets. Moreover, $n_{\text{cut}}=N/10$ in accordance with previous evidence \cite{gottwald2009implementation}. We estimated $\sigma$ of physiological noise from a given HRV series $\{RR\}$ through a closed-form methodology \cite{scarciglia2024physiological}.
Considering $\mathbb{F}$ as the magnitude of the frequency spectrum of the HRV series normalized between 0 and $2\pi$, we estimated $\Bar{c}$ as the normalized frequency corresponding to the smallest magnitude greater than zero of $\mathbb{F}$. In the case of regular dynamics, $\Bar{c}$ should be different from the main frequency of the system. Therefore, $K=\text{median}(K_c)$, considering the 100 estimates of $K$ in the interval $[\Bar{c}-0.5,\Bar{c}+0.5]$. 

\section{Experimental Data}
\subsection{Synthetic data}
We validated the proposed methodology on the following synthetic datasets: 
\begin{itemize}
    \item Pomeau-Manneville maps: a family of \emph{chaotic} maps described by the function
                        $$T_{\gamma}: [0,1]\rightarrow[0,1], \ \ \ \ T_{\gamma}(x)=\{x+x^{\gamma+1}+{\color{black}\epsilon(\sigma)}\}$$
        where $\gamma$ is a real number greater than 0, $\epsilon(\sigma)$ is a realization of a Gaussian process with mean equal 0 and standard deviation $\sigma$, and the braces denote the reduction modulo 1.
        For each $\gamma$, there exists only one $T_{\gamma}$-invariant measure $\nu_{\gamma}$, that is finite for $0<\gamma<1$, and infinite for $\gamma \geq 1$. In study, we choose $\gamma=0.2$.
    
    \item Logistic maps: a family of maps given by 
                        $$f_{\mu}: [0,1] \rightarrow [0,1], \ \ \ f_{\mu}(x)=\mu x(1-x) + {\color{black}\epsilon(\sigma)}$$
        where $\mu$ is a real-valued parameter in the interval $[0, 4]$ and $\epsilon(\sigma)$ is a realization of noise as defined above. Simulations were performed with $\mu=3.5$, in the \emph{periodic} regime, and with $\mu=3.97$, in the \emph{chaotic} regime.
\end{itemize}

We generated the dynamically perturbed series by fixing an initial condition \(x(0)\), which we exploited to produce a noise-free series of \(N=10000\) samples. Then, we considered a realization of a white Gaussian process (10000 samples) with a standard deviation equal to a percentage of the amplitude of the noise-free series, namely \(2\%, 5\%, 10\%, 15\%, 20\%\). Finally, restarting from \(x(0)\), we construct the perturbed series by adding a sample of the noise realization at each step of the map equation. For each noise level and parameter, a total of 100 realizations were obtained.
For the Logistic map, we also considered a bounce effect: each time noise leads to values out of \([0, 1]\), we put the sample back into the box \([0, 1]\) through reduction modulo 1. Such an effect is automatically verified by the reduction modulo 1 in the Pomeau-Manneville maps definition.

\subsection{Real HRV Series}

The analysis of real HRV series comprised two main datasets. 

\textbf{Dataset 1}. We analyzed long-term HRV series from 18 healthy individuals (NS, age range 20–50) gathered from the MIT-BIH Normal Sinus Rhythm Database \cite{goldberger2000physiobank} (https://physionet.org/content/nsrdb/1.0.0/), along with 23 long-term HRV series from 23 patients with atrial fibrillation (AF) gathered from the MIT-BIH Atrial Fibrillation Database \cite{mark1988bih} (https://physionet.org/content/afdb/1.0.0/). Additionally, we included long-term HRV series from 29 patients (age range 34–79) with congestive heart failure (CHF) (NYHA classes I, II, and III) gathered from the Congestive Heart Failure RR Interval Database \cite{goldberger2000physiobank} (https://physionet.org/content/chf2db/1.0.0/). While CHF series were derived from 24-hour monitoring, AF and NS series were derived from ECG series, sampled at 250Hz and 128Hz, respectively, by applying the well-known Pan–Tompkins algorithm for the identification of R-peaks. HRV series were visually inspected for physiological and algorithmic artifacts and were eventually corrected through a point-process-based software \cite{citi2012real}. To mitigate potential biases related to series length, a 10000-sample artifact-free segment was chosen for each subject.

\textbf{Dataset 2}. We analyzed long-term HRV series gathered from the "Is the normal heart rate chaotic?" dataset \cite{glass2009introduction,goldberger2000physiobank} (https://www.physionet.org/content/chaos-heart-rate/1.0.0/).
The dataset comprised 15 RR-interval time series: namely, 5 series from healthy subjects \{n1nn, n2nn, n3nn, n4nn, n5nn\}, 5 series from patients with congestive heart failure \{c1nn, c2nn, c3nn, c4nn, c5nn\}, 5 series from patients with atrial fibrillation \{a1nn, a2nn, a3nn, a4nn, a5nn\}). All of the artifact-free time series were derived from continuous ambulatory ECGs, and each time series is about 24 hours long (roughly 100,000 intervals). To the extent possible, these series contain only intervals between consecutive normal heart beats.
in order to test for non-stationary (regular or chaotic) dynamics, the analysis of the second dataset comprised following steps:
\begin{enumerate}
    \item We segmented the series into subseries comprising 5000 data points, with each subsequent subseries shifted by 2500 points from the previous one. Specifically, the initial subseries consists of the first 5000 points of the original series, followed by the second subseries formed from points 2500 to 5000 of the original series, and so forth. The final subseries consists of the last 5000 points of the original series.
    \item For each subseries, we estimated the physiological noise standard deviation \cite{scarciglia2024physiological}.
    \item We applied the test to the original series, utilizing the mean of the standard deviations of the subseries to determine the parameter $\beta$.
\end{enumerate}
\begin{figure}[!h]
\includegraphics[width=8.5cm]{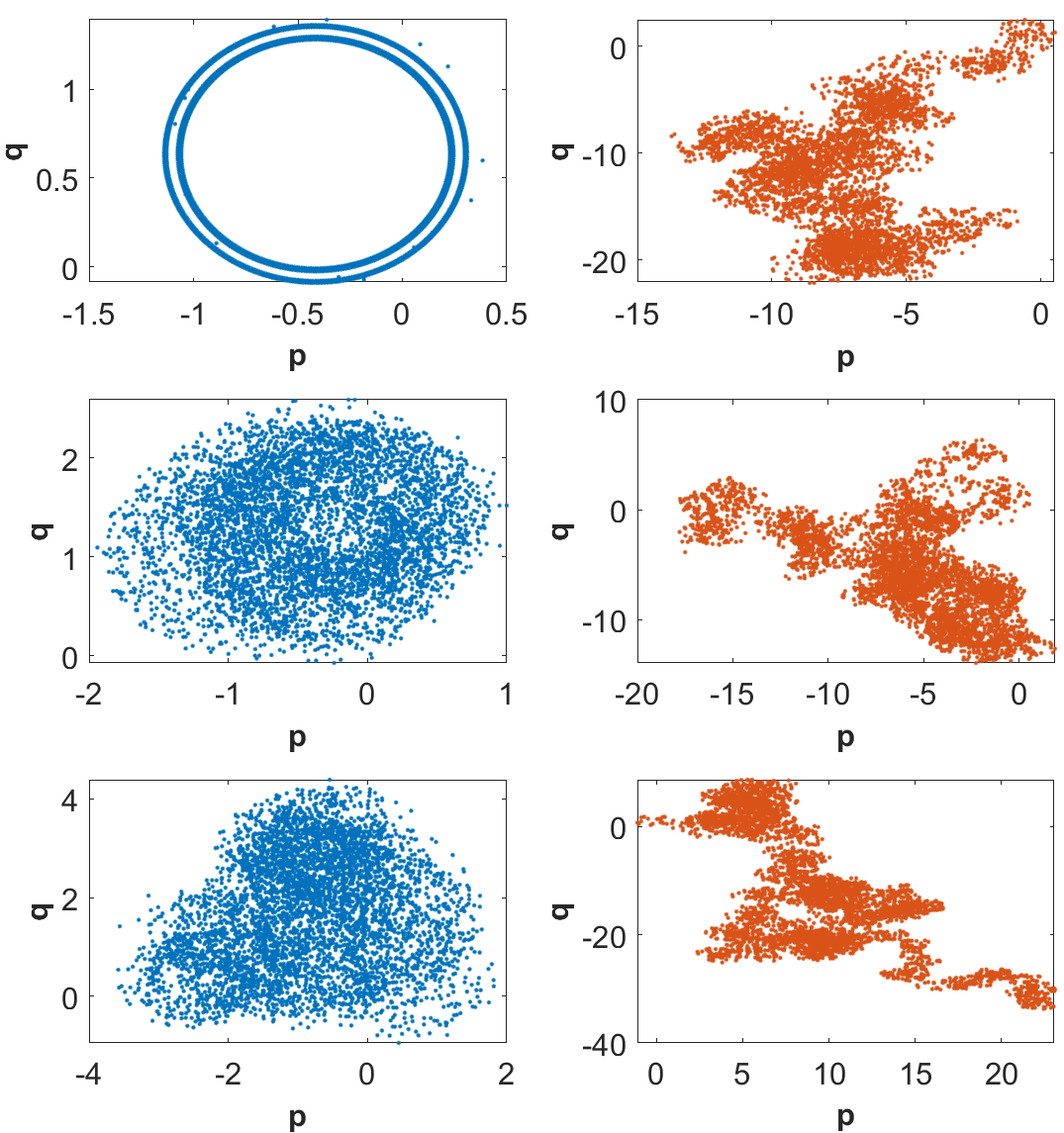}
\caption{\label{fig1} $p-q$ phase space for regular (left panels) and chaotic Logic map series (right panels) in noise-free case (first row), $\sigma$=2\% of the noise-free series range (second row), and $\sigma$=5\% of the noise-free series range (third row).}
\end{figure}

\begin{figure}[!h]
\includegraphics[width=9cm]{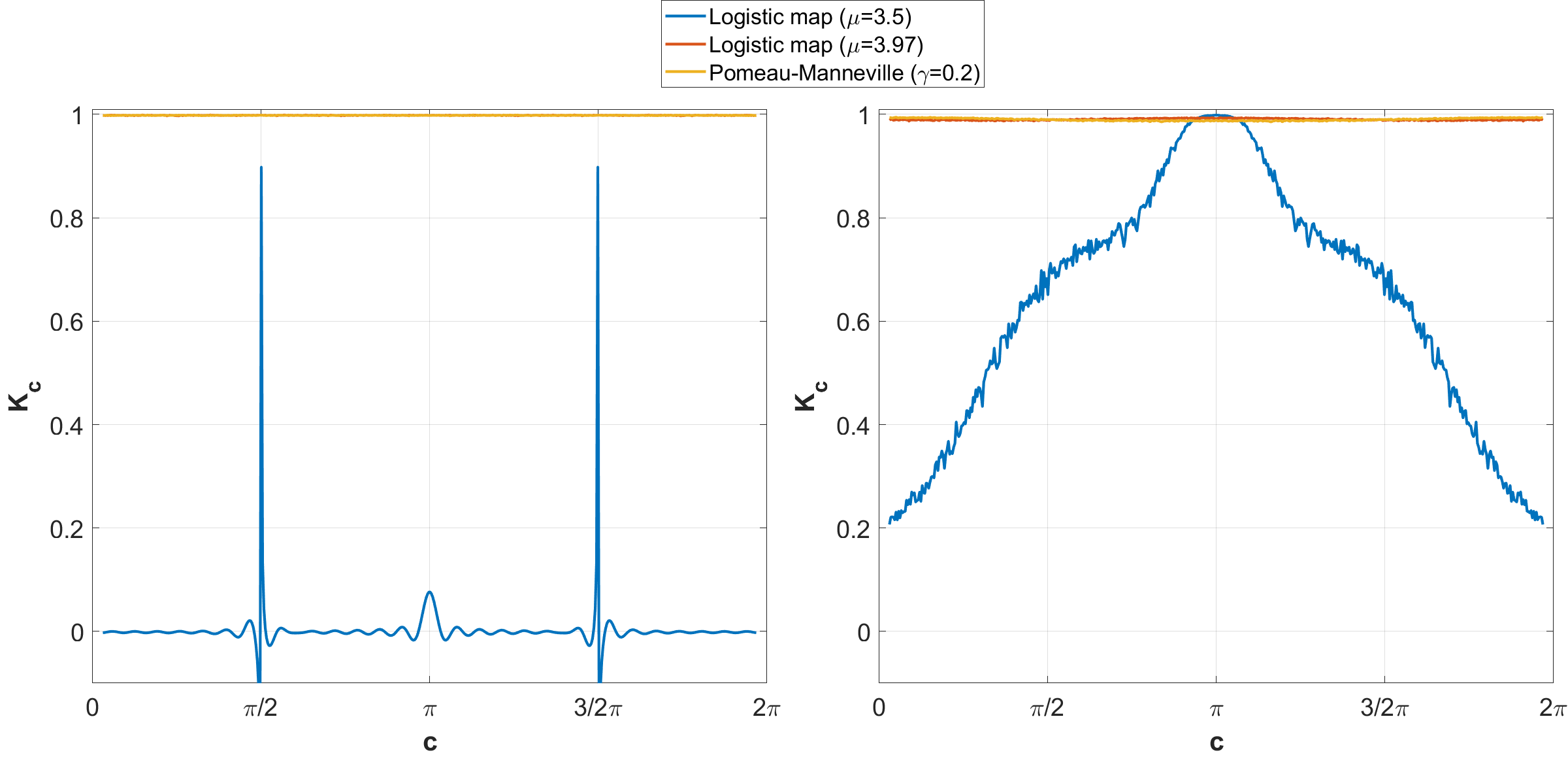}
\caption{\label{fig2} Plot of $K$ versus $c$ for the regular Logistic map (blue line), chaotic Logistic map (orange line), and Pomeau-Manneville map (dark yellow line): on the left panel is the case of no noise, on the right panel is with noise standard deviation equal to 5\% of the noise-free series range. For each map, the value of $K$ corresponding to each $c$ is the median obtained from 100 different series.}
\end{figure}

\section{Results}
\subsection{Synthetic data}

Exemplary $p-q$ phase spaces for regular and chaotic Logistic map series are illustrated in Fig. \ref{fig1}. Note that, while regular dynamics lead to bounded trajectories even in the presence of noise, chaotic dynamics lead to diffuse and unbounded trajectories.

Fig. \ref{fig2} shows exemplary plots of $K$ versus $c$ for regular and chaotic Logistic map series, as well as for Pomeau-Manneville chaotic maps in both the absence and presence of dynamical noise. Irrespective of the influence of dynamical noise, while $K<0.9$ for the vast majority of the $c$ range in the case of regular dynamics, chaotic dynamics drives $K \simeq 1$ for all $c$.

Comprehensive results in terms of $K$ statistics are presented in Fig. \ref{fig6} as boxplots, and extensively reported in Table S1 and Table S2 of the Supplementary Materials. 
Considering an empirical threshold of $K=0.9$ the largest majority of series associated with regular dynamics were with $K<0.9$, while series associated with chaotic dynamics were with $K>0.95$, throughout different noise levels.

\begin{figure}[!h]
\includegraphics[width=9cm]{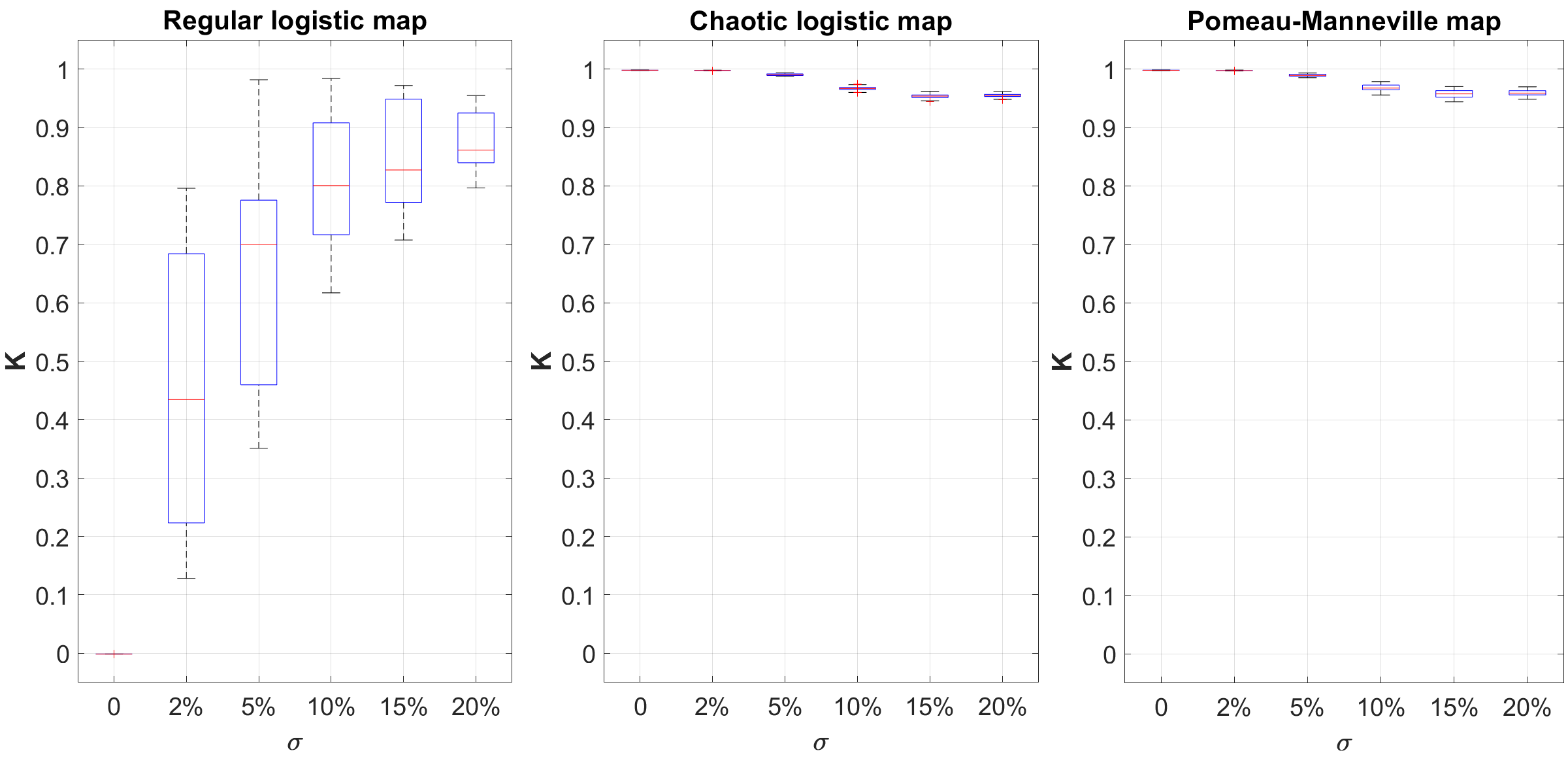}
\caption{\label{fig6} Boxplot statistics for the regular logistic maps, chaotic logistic maps, and Pomeau-Manneville maps under noise-free conditions, as well as with standard deviations of \(2\%, 5\%, 10\%, 15\%,\) and \(20\%\) noise.}
\end{figure}

\subsection{Real HRV Series}

\begin{figure}[!h]
\includegraphics[width=8.5cm]{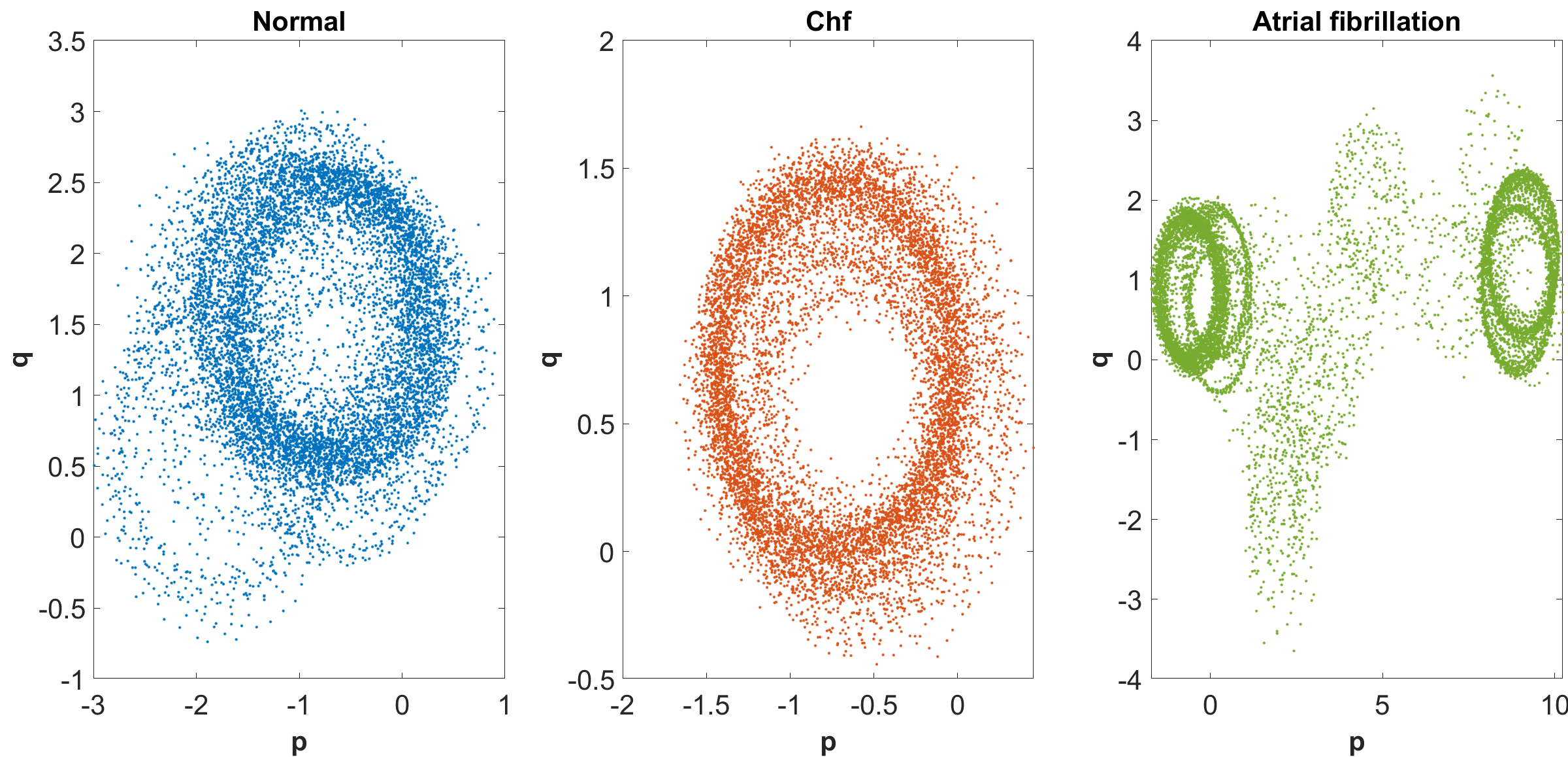}
\caption{\label{fig4} Exemplary $p-q$ phase space for an exemplary HRV series of a healthy subject (left panel), a patient with CHF (center panel), and a patient with AF (right panel).}
\end{figure}

Fig. \ref{fig4} shows the $p-q$ phase space for one exemplary recording each from a healthy subject (NS), a patient with CHF, and a patient with AF from Dataset 1. It is noteworthy that bounded trajectories are observable in all three phase spaces.

\begin{figure}[!h]
\includegraphics[width=8.5cm]{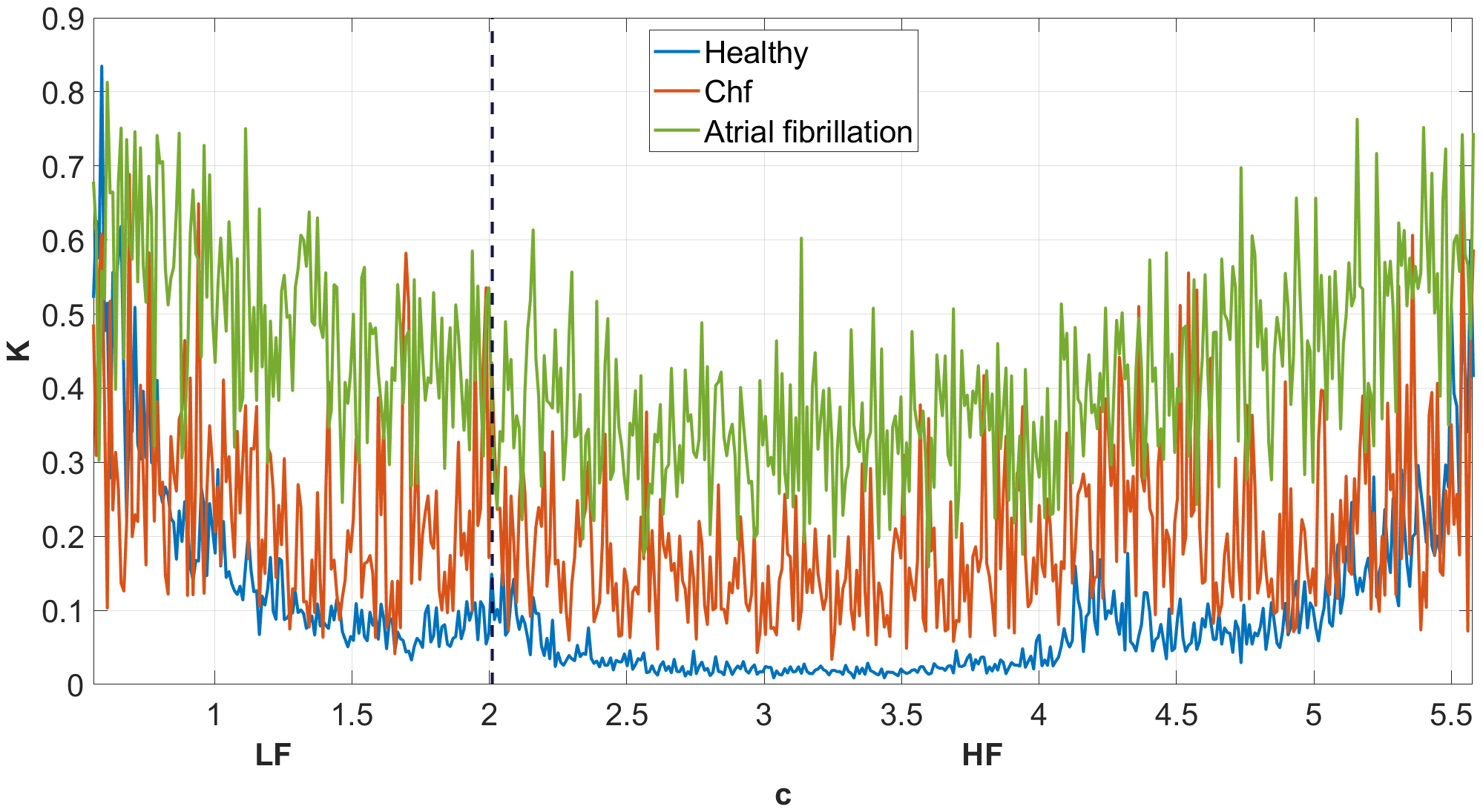}
\caption{\label{fig5} Plot of $K$ versus $c$ for n1nn (blue line), c1nn (orange line), and a1nn (green line) series. Vertical dotted line separates LF and HF frequency bands.}
\end{figure}

 Fig. \ref{fig5} illustrates the trends of $K$ versus $c$ for a subject of each experimental group from Dataset 2. While AF tends toward higher $K$ values for almost all $c$ values, the trends from all group recordings are significantly below 0.9, indicating regular dynamics despite HRV series being influenced by dynamical noise \cite{scarciglia2024physiological}.

On Dataset 1, comprehensive results regarding $K$ statistics are shown in Table \ref{tab:tableMIT} as median $\pm$ median absolute deviation intervals. While patients with AF exhibit higher physiological noise compared to CHF patients and NS, all recordings from all groups display $K<0.9$, indicating that HRV series are outputs of non-chaotic, i.e., regular, dynamical systems driven by dynamical noise. To investigate whether sympathovagal activity within the LF band and vagal activity in the HF band exhibit similar dynamics, we calculated $K$ as the median values of all $c$ values within each band. Even in these cases, for each band, $K<0.9$, further confirming that sympathovagal activity on cardiovascular control dynamics arise from non-chaotic, regular systems.

\begin{table}[!h]
\caption{\label{tab:tableMIT} Noise $\sigma$ (first column) and $K$ statistics calculated on Dataset 1 considering dynamics in the LF band (second column) and HF band (third column) computed for each group.}
\begin{ruledtabular}
\begin{tabular}{|c|cccc|}
  & Noise $\sigma$ & $K (c\in LF)$ & $K (c\in HF)$ &\\ \hline
NS & $0.023 \pm 0.009$ & $0.015 \pm 0.008$ & $0.004 \pm 0.003$ & \\ \hline
CHF & $0.015 \pm 0.004$ & $0.000 \pm 0.005$ & $0.000 \pm 0.003$ & \\   \hline
AF & $0.059 \pm 0.044$ & $0.143 \pm 0.110$ & $0.072 \pm 0.081$ &\\
\end{tabular}
\end{ruledtabular}
\end{table}

Similar results are observed with Dataset 2. $K$ statistics for each individual recording are provided in Table \ref{tab:table4} for both the LF and HF bands. Once more, despite patients with AF displaying higher physiological noise in comparison to CHF patients and NS, all recordings from all groups demonstrate $K<0.9$, suggesting that HRV series originate from regular dynamical systems influenced by dynamical noise.

\begin{table}[!h]
\caption{Noise $\sigma$ (first column) and $K$ statistics calculated on Dataset 2 considering dynamics in the LF band (second column) and HF band (third column) computed for individual recording.\label{tab:table4}}
\begin{ruledtabular}
\begin{tabular}{|c|cccc|}
  & Noise $\sigma$ & $K (c\in LF)$ & $K (c\in HF)$ &\\ \hline
n1nn & $0.031 \pm 0.009$ & 0.115 & 0.043& \\
n2nn & $0.022 \pm 0.004$ & 0.038 & 0.016& \\
n3nn & $0.029 \pm 0.005$ & 0.083 & 0.018 &\\
n4nn & $0.039 \pm 0.014$ & 0.135 & 0.039 &\\
n5nn & $0.025\pm 0.003$ & 0.036 & 0.007 &\\ \hline
c1nn & $0.012 \pm 0.002$ & 0.212 & 0.171 &\\
c2nn & $0.008 \pm 0.001$ & 0.490 & 0.403 &\\
c3nn & $0.012 \pm 0.001$ & 0.011 & 0.013 &\\
c4nn & $0.011 \pm 0.002$ & 0.035 & 0.017 &\\
c5nn & $0.011 \pm 0.004$ & 0.039 & 0.015 &\\ \hline
 a1nn & $0.166 \pm 0.014$ & 0.507 & 0.397 &\\
 a2nn & $0.113 \pm 0.012$ & 0.405 & 0.460 &\\
 a3nn & $0.181 \pm 0.026$ & 0.523 & 0.452 &\\
 a4nn & $0.112 \pm 0.018$ & 0.514 & 0.448 &\\
 a5nn & $0.107 \pm 0.018$ & 0.401 & 0.333 &\\
\end{tabular}
\end{ruledtabular}
\end{table}

\section{Discussion and Conclusion}
We introduced a novel methodology capable of distinguishing between chaotic and regular dynamics from experimental time series, which are assumed to stem from a dynamical system driven by noise. Taking inspiration from the 0-1 test for chaos and related endeavors from Gottwald and Melbourne \cite{gottwald20160,gottwald2009implementation}, we further tailored it specifically for assessing autonomic control-related cardiovascular dynamics. Our approach takes into consideration physiological noise, which recent studies have shown to significantly contribute to heartbeat dynamics \cite{scarciglia2024physiological}. Additionally, our methodology utilizes the Inverse Gaussian probability density function to quantify boundedness in phase space dynamics. This choice is motivated by the fact that the Inverse Gaussian characterizes the generative mechanisms underlying heartbeat dynamics. In fact, it accounts for the interbeat interval probability structure generated by integrate-and-fire models driven by white Gaussian noise with drift \cite{barbieri2005point,valenza2013point,valenza2018measures}. While the proposed methodology assigns a theoretical value of $K=1$ to chaotic dynamics and $K=0$ to regular dynamics, a practical estimation algorithm requires a threshold to determine the behavior of the series from a given time series. We adopted a threshold of $K=0.9$ as the best compromise between theoretical values to define chaotic dynamics above the threshold. Although this threshold may not be optimal, results from real heartbeat dynamics series were consistently associated with  $K<<0.9$, unequivocally indicating regular dynamics in the cardiovascular system.

We validated our methodology using synthetic data generated at various levels of dynamical noise. While noise does not affect the assessment of chaotic dynamics — all series generated by synthetic chaotic systems were identified with 100\% accuracy— levels of dynamical noise exceeding 10\% slightly influenced the assessment of regular dynamics. For example, 70\% of the series generated from regular Logistic maps with as much as 20\% dynamical noise were correctly identified as regular dynamics.

Experimental results from both Datasets 1 and 2 unequivocally demonstrate that heartbeat dynamics, as observed through HRV series, originate from regular systems influenced by dynamical noise. This conclusion is consistent with previous speculations \cite{glass2009introduction}, which evaluated various research studies on dynamical systems and cardiac dynamics. Our proposed methodology provides a quantitative assessment of these dynamics while accounting for the system's dynamical noise. The analysis of sympathovagal or vagally-driven autonomic control through $K$ analysis in the LF and HF bands of heartbeat dynamics did not affect the identification of regular dynamics within the cardiovascular system. This suggests that the nonlinearity of the cardiovascular system \cite{sunagawa1998dynamic,sassi2015advances} may not lead to chaotic behavior.

Understanding the regularity or chaoticity of heartbeat dynamics is crucial for cardiac modeling and the development of tailored signal processing methods for HRV series. Although cardiac pathologies such as AF and CHF do not significantly alter heartbeat dynamics in terms of regularity, our findings shed light on the regularity of AF dynamics, which may appear macroscopically irregular.

Future research will focus on applying the proposed methodological framework to neurophysiological time series, including electroencephalography, functional magnetic imaging, and functional near-infrared spectroscopy series.

\section{Supplementary Materials}
Extensive results of the synthetic data analysis are reported in the tables of the Supplementary Materials.

\section{Conflict of Interest Statement}
The authors declare no conflicts to disclose.

\section{Ethics Approval Statement }
This study was approved by the committee of bioethics of the University of Pisa with review n. 19/2021.

\section{Data Availability Statement}
All data is publicly available online. \textbf{Dataset 1}: the MIT-BIH Normal Sinus Rhythm Database \cite{goldberger2000physiobank} (https://physionet.org/content/nsrdb/1.0.0/); the MIT-BIH Atrial Fibrillation Database \cite{mark1988bih} (https://physionet.org/content/afdb/1.0.0/); the Congestive Heart Failure RR Interval Database \cite{goldberger2000physiobank} (https://physionet.org/content/chf2db/1.0.0/). 
\textbf{Dataset 2}: "Is the normal heart rate chaotic?" dataset \cite{glass2009introduction,goldberger2000physiobank} (https://www.physionet.org/content/chaos-heart-rate/1.0.0/).

\section{Author Contributions}
CB and GV designed the study. MB devised the estimation algorithm and developed the software. MB and AS analyzed the data. All authors interpreted the data and experimental results. MB wrote the manuscript. All authors revised and approved the final manuscript.

\nocite{*}
\bibliography{bibliography}

\providecommand{\noopsort}[1]{}\providecommand{\singleletter}[1]{#1}%
\begin{thebibliography}{56}%
\makeatletter
\providecommand \@ifxundefined [1]{%
 \@ifx{#1\undefined}
}%
\providecommand \@ifnum [1]{%
 \ifnum #1\expandafter \@firstoftwo
 \else \expandafter \@secondoftwo
 \fi
}%
\providecommand \@ifx [1]{%
 \ifx #1\expandafter \@firstoftwo
 \else \expandafter \@secondoftwo
 \fi
}%
\providecommand \natexlab [1]{#1}%
\providecommand \enquote  [1]{``#1''}%
\providecommand \bibnamefont  [1]{#1}%
\providecommand \bibfnamefont [1]{#1}%
\providecommand \citenamefont [1]{#1}%
\providecommand \href@noop [0]{\@secondoftwo}%
\providecommand \href [0]{\begingroup \@sanitize@url \@href}%
\providecommand \@href[1]{\@@startlink{#1}\@@href}%
\providecommand \@@href[1]{\endgroup#1\@@endlink}%
\providecommand \@sanitize@url [0]{\catcode `\\12\catcode `\$12\catcode `\&12\catcode `\#12\catcode `\^12\catcode `\_12\catcode `\%12\relax}%
\providecommand \@@startlink[1]{}%
\providecommand \@@endlink[0]{}%
\providecommand \url  [0]{\begingroup\@sanitize@url \@url }%
\providecommand \@url [1]{\endgroup\@href {#1}{\urlprefix }}%
\providecommand \urlprefix  [0]{URL }%
\providecommand \Eprint [0]{\href }%
\providecommand \doibase [0]{http://dx.doi.org/}%
\providecommand \selectlanguage [0]{\@gobble}%
\providecommand \bibinfo  [0]{\@secondoftwo}%
\providecommand \bibfield  [0]{\@secondoftwo}%
\providecommand \translation [1]{[#1]}%
\providecommand \BibitemOpen [0]{}%
\providecommand \bibitemStop [0]{}%
\providecommand \bibitemNoStop [0]{.\EOS\space}%
\providecommand \EOS [0]{\spacefactor3000\relax}%
\providecommand \BibitemShut  [1]{\csname bibitem#1\endcsname}%
\let\auto@bib@innerbib\@empty
\bibitem [{\citenamefont {Akselrod}\ \emph {et~al.}(1981)\citenamefont {Akselrod}, \citenamefont {Gordon}, \citenamefont {Ubel}, \citenamefont {Shannon}, \citenamefont {Berger},\ and\ \citenamefont {Cohen}}]{akselrod1981power}%
  \BibitemOpen
  \bibfield  {author} {\bibinfo {author} {\bibfnamefont {S.}~\bibnamefont {Akselrod}}, \bibinfo {author} {\bibfnamefont {D.}~\bibnamefont {Gordon}}, \bibinfo {author} {\bibfnamefont {F.~A.}\ \bibnamefont {Ubel}}, \bibinfo {author} {\bibfnamefont {D.~C.}\ \bibnamefont {Shannon}}, \bibinfo {author} {\bibfnamefont {A.~C.}\ \bibnamefont {Berger}}, \ and\ \bibinfo {author} {\bibfnamefont {R.~J.}\ \bibnamefont {Cohen}},\ }\bibfield  {title} {\enquote {\bibinfo {title} {Power spectrum analysis of heart rate fluctuation: a quantitative probe of beat-to-beat cardiovascular control},}\ }\href@noop {} {\bibfield  {journal} {\bibinfo  {journal} {science}\ }\textbf {\bibinfo {volume} {213}},\ \bibinfo {pages} {220--222} (\bibinfo {year} {1981})}\BibitemShut {NoStop}%
\bibitem [{\citenamefont {Saul}\ and\ \citenamefont {Valenza}(2021)}]{saul2021heart}%
  \BibitemOpen
  \bibfield  {author} {\bibinfo {author} {\bibfnamefont {J.~P.}\ \bibnamefont {Saul}}\ and\ \bibinfo {author} {\bibfnamefont {G.}~\bibnamefont {Valenza}},\ }\bibfield  {title} {\enquote {\bibinfo {title} {Heart rate variability and the dawn of complex physiological signal analysis: methodological and clinical perspectives},}\ }\href@noop {} {\bibfield  {journal} {\bibinfo  {journal} {Philosophical Transactions of the Royal Society A}\ }\textbf {\bibinfo {volume} {379}},\ \bibinfo {pages} {20200255} (\bibinfo {year} {2021})}\BibitemShut {NoStop}%
\bibitem [{\citenamefont {Rajendra~Acharya}\ \emph {et~al.}(2006)\citenamefont {Rajendra~Acharya}, \citenamefont {Paul~Joseph}, \citenamefont {Kannathal}, \citenamefont {Lim},\ and\ \citenamefont {Suri}}]{rajendra2006heart}%
  \BibitemOpen
  \bibfield  {author} {\bibinfo {author} {\bibfnamefont {U.}~\bibnamefont {Rajendra~Acharya}}, \bibinfo {author} {\bibfnamefont {K.}~\bibnamefont {Paul~Joseph}}, \bibinfo {author} {\bibfnamefont {N.}~\bibnamefont {Kannathal}}, \bibinfo {author} {\bibfnamefont {C.~M.}\ \bibnamefont {Lim}}, \ and\ \bibinfo {author} {\bibfnamefont {J.~S.}\ \bibnamefont {Suri}},\ }\bibfield  {title} {\enquote {\bibinfo {title} {Heart rate variability: a review},}\ }\href@noop {} {\bibfield  {journal} {\bibinfo  {journal} {Medical and biological engineering and computing}\ }\textbf {\bibinfo {volume} {44}},\ \bibinfo {pages} {1031--1051} (\bibinfo {year} {2006})}\BibitemShut {NoStop}%
\bibitem [{\citenamefont {Stauss}(2003)}]{stauss2003heart}%
  \BibitemOpen
  \bibfield  {author} {\bibinfo {author} {\bibfnamefont {H.~M.}\ \bibnamefont {Stauss}},\ }\bibfield  {title} {\enquote {\bibinfo {title} {Heart rate variability},}\ }\href@noop {} {\bibfield  {journal} {\bibinfo  {journal} {American Journal of Physiology-Regulatory, Integrative and Comparative Physiology}\ }\textbf {\bibinfo {volume} {285}},\ \bibinfo {pages} {R927--R931} (\bibinfo {year} {2003})}\BibitemShut {NoStop}%
\bibitem [{\citenamefont {Billman}(2011)}]{billman2011heart}%
  \BibitemOpen
  \bibfield  {author} {\bibinfo {author} {\bibfnamefont {G.~E.}\ \bibnamefont {Billman}},\ }\bibfield  {title} {\enquote {\bibinfo {title} {Heart rate variability--a historical perspective},}\ }\href@noop {} {\bibfield  {journal} {\bibinfo  {journal} {Frontiers in physiology}\ }\textbf {\bibinfo {volume} {2}},\ \bibinfo {pages} {86} (\bibinfo {year} {2011})}\BibitemShut {NoStop}%
\bibitem [{\citenamefont {Cygankiewicz}\ and\ \citenamefont {Zareba}(2013)}]{cygankiewicz2013heart}%
  \BibitemOpen
  \bibfield  {author} {\bibinfo {author} {\bibfnamefont {I.}~\bibnamefont {Cygankiewicz}}\ and\ \bibinfo {author} {\bibfnamefont {W.}~\bibnamefont {Zareba}},\ }\bibfield  {title} {\enquote {\bibinfo {title} {Heart rate variability},}\ }\href@noop {} {\bibfield  {journal} {\bibinfo  {journal} {Handbook of clinical neurology}\ }\textbf {\bibinfo {volume} {117}},\ \bibinfo {pages} {379--393} (\bibinfo {year} {2013})}\BibitemShut {NoStop}%
\bibitem [{\citenamefont {van Ravenswaaij-Arts}\ \emph {et~al.}(1993)\citenamefont {van Ravenswaaij-Arts}, \citenamefont {Kollee}, \citenamefont {Hopman}, \citenamefont {Stoelinga},\ and\ \citenamefont {van Geijn}}]{van1993heart}%
  \BibitemOpen
  \bibfield  {author} {\bibinfo {author} {\bibfnamefont {C.~M.}\ \bibnamefont {van Ravenswaaij-Arts}}, \bibinfo {author} {\bibfnamefont {L.~A.}\ \bibnamefont {Kollee}}, \bibinfo {author} {\bibfnamefont {J.~C.}\ \bibnamefont {Hopman}}, \bibinfo {author} {\bibfnamefont {G.~B.}\ \bibnamefont {Stoelinga}}, \ and\ \bibinfo {author} {\bibfnamefont {H.~P.}\ \bibnamefont {van Geijn}},\ }\bibfield  {title} {\enquote {\bibinfo {title} {Heart rate variability},}\ }\href@noop {} {\bibfield  {journal} {\bibinfo  {journal} {Annals of internal medicine}\ }\textbf {\bibinfo {volume} {118}},\ \bibinfo {pages} {436--447} (\bibinfo {year} {1993})}\BibitemShut {NoStop}%
\bibitem [{\citenamefont {Malik}(1998)}]{malik1998heart}%
  \BibitemOpen
  \bibfield  {author} {\bibinfo {author} {\bibfnamefont {M.}~\bibnamefont {Malik}},\ }\bibfield  {title} {\enquote {\bibinfo {title} {Heart rate variability},}\ }\href@noop {} {\bibfield  {journal} {\bibinfo  {journal} {Current opinion in cardiology}\ }\textbf {\bibinfo {volume} {13}},\ \bibinfo {pages} {36--44} (\bibinfo {year} {1998})}\BibitemShut {NoStop}%
\bibitem [{\citenamefont {De~Boer}, \citenamefont {Karemaker},\ and\ \citenamefont {Strackee}(1985{\natexlab{a}})}]{de1985relationships}%
  \BibitemOpen
  \bibfield  {author} {\bibinfo {author} {\bibfnamefont {R.}~\bibnamefont {De~Boer}}, \bibinfo {author} {\bibfnamefont {J.}~\bibnamefont {Karemaker}}, \ and\ \bibinfo {author} {\bibfnamefont {J.}~\bibnamefont {Strackee}},\ }\bibfield  {title} {\enquote {\bibinfo {title} {Relationships between short-term blood-pressure fluctuations and heart-rate variability in resting subjects ii: a simple model},}\ }\href@noop {} {\bibfield  {journal} {\bibinfo  {journal} {Medical and Biological Engineering and Computing}\ }\textbf {\bibinfo {volume} {23}},\ \bibinfo {pages} {359--364} (\bibinfo {year} {1985}{\natexlab{a}})}\BibitemShut {NoStop}%
\bibitem [{\citenamefont {Saul}\ \emph {et~al.}(1989)\citenamefont {Saul}, \citenamefont {Berger}, \citenamefont {Chen},\ and\ \citenamefont {Cohen}}]{saul1989transfer}%
  \BibitemOpen
  \bibfield  {author} {\bibinfo {author} {\bibfnamefont {J.~P.}\ \bibnamefont {Saul}}, \bibinfo {author} {\bibfnamefont {R.~D.}\ \bibnamefont {Berger}}, \bibinfo {author} {\bibfnamefont {M.}~\bibnamefont {Chen}}, \ and\ \bibinfo {author} {\bibfnamefont {R.~J.}\ \bibnamefont {Cohen}},\ }\bibfield  {title} {\enquote {\bibinfo {title} {Transfer function analysis of autonomic regulation. ii. respiratory sinus arrhythmia},}\ }\href@noop {} {\bibfield  {journal} {\bibinfo  {journal} {American Journal of Physiology-Heart and Circulatory Physiology}\ }\textbf {\bibinfo {volume} {256}},\ \bibinfo {pages} {H153--H161} (\bibinfo {year} {1989})}\BibitemShut {NoStop}%
\bibitem [{\citenamefont {Saul}(1991)}]{saul1991cardiorespiratory}%
  \BibitemOpen
  \bibfield  {author} {\bibinfo {author} {\bibfnamefont {J.}~\bibnamefont {Saul}},\ }\bibfield  {title} {\enquote {\bibinfo {title} {Cardiorespiratory variability: fractals, white noise, nonlinear oscillators, and linear modeling. what’s to be learned},}\ }in\ \href@noop {} {\emph {\bibinfo {booktitle} {Rhythms in Physiological Systems: Proceedings of the International Symposium at Schlo{\ss} Elmau, Bavaria, October 22--25, 1990}}}\ (\bibinfo {organization} {Springer},\ \bibinfo {year} {1991})\ pp.\ \bibinfo {pages} {115--126}\BibitemShut {NoStop}%
\bibitem [{\citenamefont {Saul}, \citenamefont {Kaplan},\ and\ \citenamefont {Kitney}(1988)}]{saul1988nonlinear}%
  \BibitemOpen
  \bibfield  {author} {\bibinfo {author} {\bibfnamefont {J.}~\bibnamefont {Saul}}, \bibinfo {author} {\bibfnamefont {D.}~\bibnamefont {Kaplan}}, \ and\ \bibinfo {author} {\bibfnamefont {R.}~\bibnamefont {Kitney}},\ }\bibfield  {title} {\enquote {\bibinfo {title} {Nonlinear interactions between respiration and heart rate: A phenomenon common to multiple physiologic states},}\ }\href@noop {} {\bibfield  {journal} {\bibinfo  {journal} {Computers in Cardiology}\ }\textbf {\bibinfo {volume} {15}},\ \bibinfo {pages} {299--302} (\bibinfo {year} {1988})}\BibitemShut {NoStop}%
\bibitem [{\citenamefont {Porta}\ \emph {et~al.}(2002)\citenamefont {Porta}, \citenamefont {Furlan}, \citenamefont {Rimoldi}, \citenamefont {Pagani}, \citenamefont {Malliani},\ and\ \citenamefont {Van De~Borne}}]{porta2002quantifying}%
  \BibitemOpen
  \bibfield  {author} {\bibinfo {author} {\bibfnamefont {A.}~\bibnamefont {Porta}}, \bibinfo {author} {\bibfnamefont {R.}~\bibnamefont {Furlan}}, \bibinfo {author} {\bibfnamefont {O.}~\bibnamefont {Rimoldi}}, \bibinfo {author} {\bibfnamefont {M.}~\bibnamefont {Pagani}}, \bibinfo {author} {\bibfnamefont {A.}~\bibnamefont {Malliani}}, \ and\ \bibinfo {author} {\bibfnamefont {P.}~\bibnamefont {Van De~Borne}},\ }\bibfield  {title} {\enquote {\bibinfo {title} {Quantifying the strength of the linear causal coupling in closed loop interacting cardiovascular variability signals},}\ }\href@noop {} {\bibfield  {journal} {\bibinfo  {journal} {Biological cybernetics}\ }\textbf {\bibinfo {volume} {86}},\ \bibinfo {pages} {241--251} (\bibinfo {year} {2002})}\BibitemShut {NoStop}%
\bibitem [{\citenamefont {Madwed}\ \emph {et~al.}(1989)\citenamefont {Madwed}, \citenamefont {Albrecht}, \citenamefont {Mark},\ and\ \citenamefont {Cohen}}]{madwed1989low}%
  \BibitemOpen
  \bibfield  {author} {\bibinfo {author} {\bibfnamefont {J.~B.}\ \bibnamefont {Madwed}}, \bibinfo {author} {\bibfnamefont {P.}~\bibnamefont {Albrecht}}, \bibinfo {author} {\bibfnamefont {R.~G.}\ \bibnamefont {Mark}}, \ and\ \bibinfo {author} {\bibfnamefont {R.~J.}\ \bibnamefont {Cohen}},\ }\bibfield  {title} {\enquote {\bibinfo {title} {Low-frequency oscillations in arterial pressure and heart rate: a simple computer model},}\ }\href@noop {} {\bibfield  {journal} {\bibinfo  {journal} {American Journal of Physiology-Heart and Circulatory Physiology}\ }\textbf {\bibinfo {volume} {256}},\ \bibinfo {pages} {H1573--H1579} (\bibinfo {year} {1989})}\BibitemShut {NoStop}%
\bibitem [{\citenamefont {Madwed}\ and\ \citenamefont {Cohen}(1991)}]{madwed1991heart}%
  \BibitemOpen
  \bibfield  {author} {\bibinfo {author} {\bibfnamefont {J.~B.}\ \bibnamefont {Madwed}}\ and\ \bibinfo {author} {\bibfnamefont {R.~J.}\ \bibnamefont {Cohen}},\ }\bibfield  {title} {\enquote {\bibinfo {title} {Heart rate response to hemorrhage-induced 0.05-hz oscillations in arterial pressure in conscious dogs},}\ }\href@noop {} {\bibfield  {journal} {\bibinfo  {journal} {American Journal of Physiology-Heart and Circulatory Physiology}\ }\textbf {\bibinfo {volume} {260}},\ \bibinfo {pages} {H1248--H1253} (\bibinfo {year} {1991})}\BibitemShut {NoStop}%
\bibitem [{\citenamefont {Baselli}\ \emph {et~al.}(1994)\citenamefont {Baselli}, \citenamefont {Cerutti}, \citenamefont {Badilini}, \citenamefont {Biancardi}, \citenamefont {Porta}, \citenamefont {Pagani}, \citenamefont {Lombardi}, \citenamefont {Rimoldi}, \citenamefont {Furlan},\ and\ \citenamefont {Malliani}}]{baselli1994model}%
  \BibitemOpen
  \bibfield  {author} {\bibinfo {author} {\bibfnamefont {G.}~\bibnamefont {Baselli}}, \bibinfo {author} {\bibfnamefont {S.}~\bibnamefont {Cerutti}}, \bibinfo {author} {\bibfnamefont {F.}~\bibnamefont {Badilini}}, \bibinfo {author} {\bibfnamefont {L.}~\bibnamefont {Biancardi}}, \bibinfo {author} {\bibfnamefont {A.}~\bibnamefont {Porta}}, \bibinfo {author} {\bibfnamefont {M.}~\bibnamefont {Pagani}}, \bibinfo {author} {\bibfnamefont {F.}~\bibnamefont {Lombardi}}, \bibinfo {author} {\bibfnamefont {O.}~\bibnamefont {Rimoldi}}, \bibinfo {author} {\bibfnamefont {R.}~\bibnamefont {Furlan}}, \ and\ \bibinfo {author} {\bibfnamefont {A.}~\bibnamefont {Malliani}},\ }\bibfield  {title} {\enquote {\bibinfo {title} {Model for the assessment of heart period and arterial pressure variability interactions and of respiration influences},}\ }\href@noop {} {\bibfield  {journal} {\bibinfo  {journal} {Medical and Biological Engineering and Computing}\ }\textbf {\bibinfo {volume} {32}},\ \bibinfo {pages} {143--152} (\bibinfo {year}
  {1994})}\BibitemShut {NoStop}%
\bibitem [{\citenamefont {Porta}\ \emph {et~al.}(2000)\citenamefont {Porta}, \citenamefont {Baselli}, \citenamefont {Rimoldi}, \citenamefont {Malliani},\ and\ \citenamefont {Pagani}}]{porta2000assessing}%
  \BibitemOpen
  \bibfield  {author} {\bibinfo {author} {\bibfnamefont {A.}~\bibnamefont {Porta}}, \bibinfo {author} {\bibfnamefont {G.}~\bibnamefont {Baselli}}, \bibinfo {author} {\bibfnamefont {O.}~\bibnamefont {Rimoldi}}, \bibinfo {author} {\bibfnamefont {A.}~\bibnamefont {Malliani}}, \ and\ \bibinfo {author} {\bibfnamefont {M.}~\bibnamefont {Pagani}},\ }\bibfield  {title} {\enquote {\bibinfo {title} {Assessing baroreflex gain from spontaneous variability in conscious dogs: role of causality and respiration},}\ }\href@noop {} {\bibfield  {journal} {\bibinfo  {journal} {American Journal of Physiology-Heart and Circulatory Physiology}\ }\textbf {\bibinfo {volume} {279}},\ \bibinfo {pages} {H2558--H2567} (\bibinfo {year} {2000})}\BibitemShut {NoStop}%
\bibitem [{\citenamefont {Valenza}\ \emph {et~al.}(2018)\citenamefont {Valenza}, \citenamefont {Citi}, \citenamefont {Saul},\ and\ \citenamefont {Barbieri}}]{valenza2018measures}%
  \BibitemOpen
  \bibfield  {author} {\bibinfo {author} {\bibfnamefont {G.}~\bibnamefont {Valenza}}, \bibinfo {author} {\bibfnamefont {L.}~\bibnamefont {Citi}}, \bibinfo {author} {\bibfnamefont {J.~P.}\ \bibnamefont {Saul}}, \ and\ \bibinfo {author} {\bibfnamefont {R.}~\bibnamefont {Barbieri}},\ }\bibfield  {title} {\enquote {\bibinfo {title} {Measures of sympathetic and parasympathetic autonomic outflow from heartbeat dynamics},}\ }\href@noop {} {\bibfield  {journal} {\bibinfo  {journal} {Journal of applied physiology}\ }\textbf {\bibinfo {volume} {125}},\ \bibinfo {pages} {19--39} (\bibinfo {year} {2018})}\BibitemShut {NoStop}%
\bibitem [{\citenamefont {Sunagawa}, \citenamefont {Kawada},\ and\ \citenamefont {Nakahara}(1998)}]{sunagawa1998dynamic}%
  \BibitemOpen
  \bibfield  {author} {\bibinfo {author} {\bibfnamefont {K.}~\bibnamefont {Sunagawa}}, \bibinfo {author} {\bibfnamefont {T.}~\bibnamefont {Kawada}}, \ and\ \bibinfo {author} {\bibfnamefont {T.}~\bibnamefont {Nakahara}},\ }\bibfield  {title} {\enquote {\bibinfo {title} {Dynamic nonlinear vago-sympathetic interaction in regulating heart rate},}\ }\href@noop {} {\bibfield  {journal} {\bibinfo  {journal} {Heart and vessels}\ }\textbf {\bibinfo {volume} {13}},\ \bibinfo {pages} {157--174} (\bibinfo {year} {1998})}\BibitemShut {NoStop}%
\bibitem [{\citenamefont {Glass}(2009)}]{glass2009introduction}%
  \BibitemOpen
  \bibfield  {author} {\bibinfo {author} {\bibfnamefont {L.}~\bibnamefont {Glass}},\ }\bibfield  {title} {\enquote {\bibinfo {title} {Introduction to controversial topics in nonlinear science: Is the normal heart rate chaotic?}}\ }\href@noop {} {\bibfield  {journal} {\bibinfo  {journal} {Chaos: An Interdisciplinary Journal of Nonlinear Science}\ }\textbf {\bibinfo {volume} {19}} (\bibinfo {year} {2009})}\BibitemShut {NoStop}%
\bibitem [{\citenamefont {Spasi{\'c}}\ and\ \citenamefont {Kesi{\'c}}(2019)}]{spasic2019nonlinearity}%
  \BibitemOpen
  \bibfield  {author} {\bibinfo {author} {\bibfnamefont {S.~Z.}\ \bibnamefont {Spasi{\'c}}}\ and\ \bibinfo {author} {\bibfnamefont {S.}~\bibnamefont {Kesi{\'c}}},\ }\bibfield  {title} {\enquote {\bibinfo {title} {Nonlinearity in living systems: Theoretical and practical perspectives on metrics of physiological signal complexity},}\ }\href@noop {} {\bibfield  {journal} {\bibinfo  {journal} {Frontiers in physiology}\ }\textbf {\bibinfo {volume} {10}},\ \bibinfo {pages} {453202} (\bibinfo {year} {2019})}\BibitemShut {NoStop}%
\bibitem [{\citenamefont {Barbieri}, \citenamefont {Scilingo},\ and\ \citenamefont {Valenza}(2017)}]{barbieri2017complexity}%
  \BibitemOpen
  \bibfield  {author} {\bibinfo {author} {\bibfnamefont {R.}~\bibnamefont {Barbieri}}, \bibinfo {author} {\bibfnamefont {E.~P.}\ \bibnamefont {Scilingo}}, \ and\ \bibinfo {author} {\bibfnamefont {G.}~\bibnamefont {Valenza}},\ }\href@noop {} {\emph {\bibinfo {title} {Complexity and nonlinearity in cardiovascular signals}}}\ (\bibinfo  {publisher} {Springer},\ \bibinfo {year} {2017})\BibitemShut {NoStop}%
\bibitem [{\citenamefont {Sassi}\ \emph {et~al.}(2015)\citenamefont {Sassi}, \citenamefont {Cerutti}, \citenamefont {Lombardi}, \citenamefont {Malik}, \citenamefont {Huikuri}, \citenamefont {Peng}, \citenamefont {Schmidt}, \citenamefont {Yamamoto}, \citenamefont {Reviewers:}, \citenamefont {Gorenek} \emph {et~al.}}]{sassi2015advances}%
  \BibitemOpen
  \bibfield  {author} {\bibinfo {author} {\bibfnamefont {R.}~\bibnamefont {Sassi}}, \bibinfo {author} {\bibfnamefont {S.}~\bibnamefont {Cerutti}}, \bibinfo {author} {\bibfnamefont {F.}~\bibnamefont {Lombardi}}, \bibinfo {author} {\bibfnamefont {M.}~\bibnamefont {Malik}}, \bibinfo {author} {\bibfnamefont {H.~V.}\ \bibnamefont {Huikuri}}, \bibinfo {author} {\bibfnamefont {C.-K.}\ \bibnamefont {Peng}}, \bibinfo {author} {\bibfnamefont {G.}~\bibnamefont {Schmidt}}, \bibinfo {author} {\bibfnamefont {Y.}~\bibnamefont {Yamamoto}}, \bibinfo {author} {\bibfnamefont {D.}~\bibnamefont {Reviewers:}}, \bibinfo {author} {\bibfnamefont {B.}~\bibnamefont {Gorenek}},  \emph {et~al.},\ }\bibfield  {title} {\enquote {\bibinfo {title} {Advances in heart rate variability signal analysis: joint position statement by the e-cardiology esc working group and the european heart rhythm association co-endorsed by the asia pacific heart rhythm society},}\ }\href@noop {} {\bibfield  {journal} {\bibinfo  {journal} {Ep Europace}\ }\textbf
  {\bibinfo {volume} {17}},\ \bibinfo {pages} {1341--1353} (\bibinfo {year} {2015})}\BibitemShut {NoStop}%
\bibitem [{\citenamefont {Takens}(2006)}]{takens2006detecting}%
  \BibitemOpen
  \bibfield  {author} {\bibinfo {author} {\bibfnamefont {F.}~\bibnamefont {Takens}},\ }\bibfield  {title} {\enquote {\bibinfo {title} {Detecting strange attractors in turbulence},}\ }in\ \href@noop {} {\emph {\bibinfo {booktitle} {Dynamical Systems and Turbulence, Warwick 1980: proceedings of a symposium held at the University of Warwick 1979/80}}}\ (\bibinfo {organization} {Springer},\ \bibinfo {year} {2006})\ pp.\ \bibinfo {pages} {366--381}\BibitemShut {NoStop}%
\bibitem [{\citenamefont {Rosenstein}, \citenamefont {Collins},\ and\ \citenamefont {De~Luca}(1993)}]{rosenstein1993practical}%
  \BibitemOpen
  \bibfield  {author} {\bibinfo {author} {\bibfnamefont {M.~T.}\ \bibnamefont {Rosenstein}}, \bibinfo {author} {\bibfnamefont {J.~J.}\ \bibnamefont {Collins}}, \ and\ \bibinfo {author} {\bibfnamefont {C.~J.}\ \bibnamefont {De~Luca}},\ }\bibfield  {title} {\enquote {\bibinfo {title} {A practical method for calculating largest lyapunov exponents from small data sets},}\ }\href@noop {} {\bibfield  {journal} {\bibinfo  {journal} {Physica D: Nonlinear Phenomena}\ }\textbf {\bibinfo {volume} {65}},\ \bibinfo {pages} {117--134} (\bibinfo {year} {1993})}\BibitemShut {NoStop}%
\bibitem [{\citenamefont {Grassberger}\ and\ \citenamefont {Procaccia}(1983)}]{grassberger1983measuring}%
  \BibitemOpen
  \bibfield  {author} {\bibinfo {author} {\bibfnamefont {P.}~\bibnamefont {Grassberger}}\ and\ \bibinfo {author} {\bibfnamefont {I.}~\bibnamefont {Procaccia}},\ }\bibfield  {title} {\enquote {\bibinfo {title} {Measuring the strangeness of strange attractors},}\ }\href@noop {} {\bibfield  {journal} {\bibinfo  {journal} {Physica D: nonlinear phenomena}\ }\textbf {\bibinfo {volume} {9}},\ \bibinfo {pages} {189--208} (\bibinfo {year} {1983})}\BibitemShut {NoStop}%
\bibitem [{\citenamefont {Kantz}\ and\ \citenamefont {Schreiber}(2004)}]{kantz2004nonlinear}%
  \BibitemOpen
  \bibfield  {author} {\bibinfo {author} {\bibfnamefont {H.}~\bibnamefont {Kantz}}\ and\ \bibinfo {author} {\bibfnamefont {T.}~\bibnamefont {Schreiber}},\ }\href@noop {} {\emph {\bibinfo {title} {Nonlinear time series analysis}}},\ Vol.~\bibinfo {volume} {7}\ (\bibinfo  {publisher} {Cambridge university press},\ \bibinfo {year} {2004})\BibitemShut {NoStop}%
\bibitem [{\citenamefont {Pincus}(1991)}]{pincus1991approximate}%
  \BibitemOpen
  \bibfield  {author} {\bibinfo {author} {\bibfnamefont {S.~M.}\ \bibnamefont {Pincus}},\ }\bibfield  {title} {\enquote {\bibinfo {title} {Approximate entropy as a measure of system complexity.}}\ }\href@noop {} {\bibfield  {journal} {\bibinfo  {journal} {Proceedings of the national academy of sciences}\ }\textbf {\bibinfo {volume} {88}},\ \bibinfo {pages} {2297--2301} (\bibinfo {year} {1991})}\BibitemShut {NoStop}%
\bibitem [{\citenamefont {Pincus}\ and\ \citenamefont {Goldberger}(1994)}]{pincus1994physiological}%
  \BibitemOpen
  \bibfield  {author} {\bibinfo {author} {\bibfnamefont {S.~M.}\ \bibnamefont {Pincus}}\ and\ \bibinfo {author} {\bibfnamefont {A.~L.}\ \bibnamefont {Goldberger}},\ }\bibfield  {title} {\enquote {\bibinfo {title} {Physiological time-series analysis: what does regularity quantify?}}\ }\href@noop {} {\bibfield  {journal} {\bibinfo  {journal} {American Journal of Physiology-Heart and Circulatory Physiology}\ }\textbf {\bibinfo {volume} {266}},\ \bibinfo {pages} {H1643--H1656} (\bibinfo {year} {1994})}\BibitemShut {NoStop}%
\bibitem [{\citenamefont {Richman}\ and\ \citenamefont {Moorman}(2000)}]{richman2000physiological}%
  \BibitemOpen
  \bibfield  {author} {\bibinfo {author} {\bibfnamefont {J.~S.}\ \bibnamefont {Richman}}\ and\ \bibinfo {author} {\bibfnamefont {J.~R.}\ \bibnamefont {Moorman}},\ }\bibfield  {title} {\enquote {\bibinfo {title} {Physiological time-series analysis using approximate entropy and sample entropy},}\ }\href@noop {} {\bibfield  {journal} {\bibinfo  {journal} {American journal of physiology-heart and circulatory physiology}\ }\textbf {\bibinfo {volume} {278}},\ \bibinfo {pages} {H2039--H2049} (\bibinfo {year} {2000})}\BibitemShut {NoStop}%
\bibitem [{\citenamefont {Li}\ \emph {et~al.}(2015)\citenamefont {Li}, \citenamefont {Liu}, \citenamefont {Li}, \citenamefont {Zheng}, \citenamefont {Liu},\ and\ \citenamefont {Hou}}]{li2015assessing}%
  \BibitemOpen
  \bibfield  {author} {\bibinfo {author} {\bibfnamefont {P.}~\bibnamefont {Li}}, \bibinfo {author} {\bibfnamefont {C.}~\bibnamefont {Liu}}, \bibinfo {author} {\bibfnamefont {K.}~\bibnamefont {Li}}, \bibinfo {author} {\bibfnamefont {D.}~\bibnamefont {Zheng}}, \bibinfo {author} {\bibfnamefont {C.}~\bibnamefont {Liu}}, \ and\ \bibinfo {author} {\bibfnamefont {Y.}~\bibnamefont {Hou}},\ }\bibfield  {title} {\enquote {\bibinfo {title} {Assessing the complexity of short-term heartbeat interval series by distribution entropy},}\ }\href@noop {} {\bibfield  {journal} {\bibinfo  {journal} {Medical \& biological engineering \& computing}\ }\textbf {\bibinfo {volume} {53}},\ \bibinfo {pages} {77--87} (\bibinfo {year} {2015})}\BibitemShut {NoStop}%
\bibitem [{\citenamefont {Scarciglia}\ \emph {et~al.}(2023)\citenamefont {Scarciglia}, \citenamefont {Gini}, \citenamefont {Catrambone}, \citenamefont {Bonanno},\ and\ \citenamefont {Valenza}}]{scarciglia2023estimation}%
  \BibitemOpen
  \bibfield  {author} {\bibinfo {author} {\bibfnamefont {A.}~\bibnamefont {Scarciglia}}, \bibinfo {author} {\bibfnamefont {F.}~\bibnamefont {Gini}}, \bibinfo {author} {\bibfnamefont {V.}~\bibnamefont {Catrambone}}, \bibinfo {author} {\bibfnamefont {C.}~\bibnamefont {Bonanno}}, \ and\ \bibinfo {author} {\bibfnamefont {G.}~\bibnamefont {Valenza}},\ }\bibfield  {title} {\enquote {\bibinfo {title} {Estimation of dynamical noise power in unknown systems},}\ }\href@noop {} {\bibfield  {journal} {\bibinfo  {journal} {IEEE Signal Processing Letters}\ }\textbf {\bibinfo {volume} {30}},\ \bibinfo {pages} {234--238} (\bibinfo {year} {2023})}\BibitemShut {NoStop}%
\bibitem [{\citenamefont {Scarciglia}\ \emph {et~al.}(2024)\citenamefont {Scarciglia}, \citenamefont {Catrambone}, \citenamefont {Bonanno},\ and\ \citenamefont {Valenza}}]{scarciglia2024physiological}%
  \BibitemOpen
  \bibfield  {author} {\bibinfo {author} {\bibfnamefont {A.}~\bibnamefont {Scarciglia}}, \bibinfo {author} {\bibfnamefont {V.}~\bibnamefont {Catrambone}}, \bibinfo {author} {\bibfnamefont {C.}~\bibnamefont {Bonanno}}, \ and\ \bibinfo {author} {\bibfnamefont {G.}~\bibnamefont {Valenza}},\ }\bibfield  {title} {\enquote {\bibinfo {title} {Physiological noise: Definition, estimation, and characterization in complex biomedical signals},}\ }\href@noop {} {\bibfield  {journal} {\bibinfo  {journal} {IEEE transactions on bio-medical engineering}\ }\textbf {\bibinfo {volume} {71}},\ \bibinfo {pages} {45--55} (\bibinfo {year} {2024})}\BibitemShut {NoStop}%
\bibitem [{\citenamefont {Manor}\ \emph {et~al.}(2010)\citenamefont {Manor}, \citenamefont {Costa}, \citenamefont {Hu}, \citenamefont {Newton}, \citenamefont {Starobinets}, \citenamefont {Kang}, \citenamefont {Peng}, \citenamefont {Novak},\ and\ \citenamefont {Lipsitz}}]{manor2010physiological}%
  \BibitemOpen
  \bibfield  {author} {\bibinfo {author} {\bibfnamefont {B.}~\bibnamefont {Manor}}, \bibinfo {author} {\bibfnamefont {M.~D.}\ \bibnamefont {Costa}}, \bibinfo {author} {\bibfnamefont {K.}~\bibnamefont {Hu}}, \bibinfo {author} {\bibfnamefont {E.}~\bibnamefont {Newton}}, \bibinfo {author} {\bibfnamefont {O.}~\bibnamefont {Starobinets}}, \bibinfo {author} {\bibfnamefont {H.~G.}\ \bibnamefont {Kang}}, \bibinfo {author} {\bibfnamefont {C.}~\bibnamefont {Peng}}, \bibinfo {author} {\bibfnamefont {V.}~\bibnamefont {Novak}}, \ and\ \bibinfo {author} {\bibfnamefont {L.~A.}\ \bibnamefont {Lipsitz}},\ }\bibfield  {title} {\enquote {\bibinfo {title} {Physiological complexity and system adaptability: evidence from postural control dynamics of older adults},}\ }\href@noop {} {\bibfield  {journal} {\bibinfo  {journal} {Journal of Applied Physiology}\ }\textbf {\bibinfo {volume} {109}},\ \bibinfo {pages} {1786--1791} (\bibinfo {year} {2010})}\BibitemShut {NoStop}%
\bibitem [{\citenamefont {San~Miguel}(2023)}]{san2023frontiers}%
  \BibitemOpen
  \bibfield  {author} {\bibinfo {author} {\bibfnamefont {M.}~\bibnamefont {San~Miguel}},\ }\bibfield  {title} {\enquote {\bibinfo {title} {Frontiers in complex systems},}\ }\href@noop {} {\bibfield  {journal} {\bibinfo  {journal} {Frontiers in Complex Systems}\ }\textbf {\bibinfo {volume} {1}},\ \bibinfo {pages} {1080801} (\bibinfo {year} {2023})}\BibitemShut {NoStop}%
\bibitem [{\citenamefont {Zhang}\ \emph {et~al.}(2009)\citenamefont {Zhang}, \citenamefont {Holden}, \citenamefont {Monfredi}, \citenamefont {Boyett},\ and\ \citenamefont {Zhang}}]{zhang2009stochastic}%
  \BibitemOpen
  \bibfield  {author} {\bibinfo {author} {\bibfnamefont {J.}~\bibnamefont {Zhang}}, \bibinfo {author} {\bibfnamefont {A.}~\bibnamefont {Holden}}, \bibinfo {author} {\bibfnamefont {O.}~\bibnamefont {Monfredi}}, \bibinfo {author} {\bibfnamefont {M.~R.}\ \bibnamefont {Boyett}}, \ and\ \bibinfo {author} {\bibfnamefont {H.}~\bibnamefont {Zhang}},\ }\bibfield  {title} {\enquote {\bibinfo {title} {Stochastic vagal modulation of cardiac pacemaking may lead to erroneous identification of cardiac “chaos”},}\ }\href@noop {} {\bibfield  {journal} {\bibinfo  {journal} {Chaos: An Interdisciplinary Journal of Nonlinear Science}\ }\textbf {\bibinfo {volume} {19}} (\bibinfo {year} {2009})}\BibitemShut {NoStop}%
\bibitem [{\citenamefont {Gottwald}\ and\ \citenamefont {Melbourne}(2016)}]{gottwald20160}%
  \BibitemOpen
  \bibfield  {author} {\bibinfo {author} {\bibfnamefont {G.~A.}\ \bibnamefont {Gottwald}}\ and\ \bibinfo {author} {\bibfnamefont {I.}~\bibnamefont {Melbourne}},\ }\bibfield  {title} {\enquote {\bibinfo {title} {The 0-1 test for chaos: A review},}\ }\href@noop {} {\bibfield  {journal} {\bibinfo  {journal} {Chaos detection and predictability}\ ,\ \bibinfo {pages} {221--247}} (\bibinfo {year} {2016})}\BibitemShut {NoStop}%
\bibitem [{\citenamefont {Gottwald}\ and\ \citenamefont {Melbourne}(2009)}]{gottwald2009implementation}%
  \BibitemOpen
  \bibfield  {author} {\bibinfo {author} {\bibfnamefont {G.~A.}\ \bibnamefont {Gottwald}}\ and\ \bibinfo {author} {\bibfnamefont {I.}~\bibnamefont {Melbourne}},\ }\bibfield  {title} {\enquote {\bibinfo {title} {On the implementation of the 0--1 test for chaos},}\ }\href@noop {} {\bibfield  {journal} {\bibinfo  {journal} {SIAM Journal on Applied Dynamical Systems}\ }\textbf {\bibinfo {volume} {8}},\ \bibinfo {pages} {129--145} (\bibinfo {year} {2009})}\BibitemShut {NoStop}%
\bibitem [{\citenamefont {Valenza}\ \emph {et~al.}(2013)\citenamefont {Valenza}, \citenamefont {Citi}, \citenamefont {Scilingo},\ and\ \citenamefont {Barbieri}}]{valenza2013point}%
  \BibitemOpen
  \bibfield  {author} {\bibinfo {author} {\bibfnamefont {G.}~\bibnamefont {Valenza}}, \bibinfo {author} {\bibfnamefont {L.}~\bibnamefont {Citi}}, \bibinfo {author} {\bibfnamefont {E.~P.}\ \bibnamefont {Scilingo}}, \ and\ \bibinfo {author} {\bibfnamefont {R.}~\bibnamefont {Barbieri}},\ }\bibfield  {title} {\enquote {\bibinfo {title} {Point-process nonlinear models with laguerre and volterra expansions: Instantaneous assessment of heartbeat dynamics},}\ }\href@noop {} {\bibfield  {journal} {\bibinfo  {journal} {IEEE Transactions on Signal Processing}\ }\textbf {\bibinfo {volume} {61}},\ \bibinfo {pages} {2914--2926} (\bibinfo {year} {2013})}\BibitemShut {NoStop}%
\bibitem [{\citenamefont {Goldberger}\ \emph {et~al.}(2000)\citenamefont {Goldberger}, \citenamefont {Amaral}, \citenamefont {Glass}, \citenamefont {Hausdorff}, \citenamefont {Ivanov}, \citenamefont {Mark}, \citenamefont {Mietus}, \citenamefont {Moody}, \citenamefont {Peng},\ and\ \citenamefont {Stanley}}]{goldberger2000physiobank}%
  \BibitemOpen
  \bibfield  {author} {\bibinfo {author} {\bibfnamefont {A.~L.}\ \bibnamefont {Goldberger}}, \bibinfo {author} {\bibfnamefont {L.~A.}\ \bibnamefont {Amaral}}, \bibinfo {author} {\bibfnamefont {L.}~\bibnamefont {Glass}}, \bibinfo {author} {\bibfnamefont {J.~M.}\ \bibnamefont {Hausdorff}}, \bibinfo {author} {\bibfnamefont {P.~C.}\ \bibnamefont {Ivanov}}, \bibinfo {author} {\bibfnamefont {R.~G.}\ \bibnamefont {Mark}}, \bibinfo {author} {\bibfnamefont {J.~E.}\ \bibnamefont {Mietus}}, \bibinfo {author} {\bibfnamefont {G.~B.}\ \bibnamefont {Moody}}, \bibinfo {author} {\bibfnamefont {C.-K.}\ \bibnamefont {Peng}}, \ and\ \bibinfo {author} {\bibfnamefont {H.~E.}\ \bibnamefont {Stanley}},\ }\bibfield  {title} {\enquote {\bibinfo {title} {Physiobank, physiotoolkit, and physionet: components of a new research resource for complex physiologic signals},}\ }\href@noop {} {\bibfield  {journal} {\bibinfo  {journal} {circulation}\ }\textbf {\bibinfo {volume} {101}},\ \bibinfo {pages} {e215--e220} (\bibinfo {year}
  {2000})}\BibitemShut {NoStop}%
\bibitem [{\citenamefont {Mark}\ and\ \citenamefont {Moody}(1988)}]{mark1988bih}%
  \BibitemOpen
  \bibfield  {author} {\bibinfo {author} {\bibfnamefont {R.}~\bibnamefont {Mark}}\ and\ \bibinfo {author} {\bibfnamefont {G.}~\bibnamefont {Moody}},\ }\bibfield  {title} {\enquote {\bibinfo {title} {Mit-bih arrhythmia database directory},}\ }\href@noop {} {\bibfield  {journal} {\bibinfo  {journal} {Cambridge: Massachusetts Institute of Technology}\ } (\bibinfo {year} {1988})}\BibitemShut {NoStop}%
\bibitem [{\citenamefont {Citi}, \citenamefont {Brown},\ and\ \citenamefont {Barbieri}(2012)}]{citi2012real}%
  \BibitemOpen
  \bibfield  {author} {\bibinfo {author} {\bibfnamefont {L.}~\bibnamefont {Citi}}, \bibinfo {author} {\bibfnamefont {E.~N.}\ \bibnamefont {Brown}}, \ and\ \bibinfo {author} {\bibfnamefont {R.}~\bibnamefont {Barbieri}},\ }\bibfield  {title} {\enquote {\bibinfo {title} {A real-time automated point-process method for the detection and correction of erroneous and ectopic heartbeats},}\ }\href@noop {} {\bibfield  {journal} {\bibinfo  {journal} {IEEE transactions on biomedical engineering}\ }\textbf {\bibinfo {volume} {59}},\ \bibinfo {pages} {2828--2837} (\bibinfo {year} {2012})}\BibitemShut {NoStop}%
\bibitem [{\citenamefont {Barbieri}\ \emph {et~al.}(2005)\citenamefont {Barbieri}, \citenamefont {Matten}, \citenamefont {Alabi},\ and\ \citenamefont {Brown}}]{barbieri2005point}%
  \BibitemOpen
  \bibfield  {author} {\bibinfo {author} {\bibfnamefont {R.}~\bibnamefont {Barbieri}}, \bibinfo {author} {\bibfnamefont {E.~C.}\ \bibnamefont {Matten}}, \bibinfo {author} {\bibfnamefont {A.~A.}\ \bibnamefont {Alabi}}, \ and\ \bibinfo {author} {\bibfnamefont {E.~N.}\ \bibnamefont {Brown}},\ }\bibfield  {title} {\enquote {\bibinfo {title} {A point-process model of human heartbeat intervals: new definitions of heart rate and heart rate variability},}\ }\href@noop {} {\bibfield  {journal} {\bibinfo  {journal} {American Journal of Physiology-Heart and Circulatory Physiology}\ }\textbf {\bibinfo {volume} {288}},\ \bibinfo {pages} {H424--H435} (\bibinfo {year} {2005})}\BibitemShut {NoStop}%
\bibitem [{\citenamefont {Krum}\ \emph {et~al.}(1995)\citenamefont {Krum}, \citenamefont {Bigger~Jr}, \citenamefont {Goldsmith},\ and\ \citenamefont {Packer}}]{krum1995effect}%
  \BibitemOpen
  \bibfield  {author} {\bibinfo {author} {\bibfnamefont {H.}~\bibnamefont {Krum}}, \bibinfo {author} {\bibfnamefont {T.}~\bibnamefont {Bigger~Jr}}, \bibinfo {author} {\bibfnamefont {R.~L.}\ \bibnamefont {Goldsmith}}, \ and\ \bibinfo {author} {\bibfnamefont {M.}~\bibnamefont {Packer}},\ }\bibfield  {title} {\enquote {\bibinfo {title} {Effect of long-term digoxin therapy on autonomic function in patients with chronic heart failure},}\ }\href@noop {} {\bibfield  {journal} {\bibinfo  {journal} {Journal of the American College of Cardiology}\ }\textbf {\bibinfo {volume} {25}},\ \bibinfo {pages} {289--294} (\bibinfo {year} {1995})}\BibitemShut {NoStop}%
\bibitem [{\citenamefont {Moody}(1983)}]{moody1983new}%
  \BibitemOpen
  \bibfield  {author} {\bibinfo {author} {\bibfnamefont {G.}~\bibnamefont {Moody}},\ }\bibfield  {title} {\enquote {\bibinfo {title} {A new method for detecting atrial fibrillation using rr intervals},}\ }\href@noop {} {\bibfield  {journal} {\bibinfo  {journal} {Proc. Comput. Cardiol.}\ }\textbf {\bibinfo {volume} {10}},\ \bibinfo {pages} {227--230} (\bibinfo {year} {1983})}\BibitemShut {NoStop}%
\bibitem [{\citenamefont {Pan}\ and\ \citenamefont {Tompkins}(1985)}]{4122029}%
  \BibitemOpen
  \bibfield  {author} {\bibinfo {author} {\bibfnamefont {J.}~\bibnamefont {Pan}}\ and\ \bibinfo {author} {\bibfnamefont {W.~J.}\ \bibnamefont {Tompkins}},\ }\bibfield  {title} {\enquote {\bibinfo {title} {A real-time qrs detection algorithm},}\ }\href {\doibase 10.1109/TBME.1985.325532} {\bibfield  {journal} {\bibinfo  {journal} {IEEE Transactions on Biomedical Engineering}\ }\textbf {\bibinfo {volume} {BME-32}},\ \bibinfo {pages} {230--236} (\bibinfo {year} {1985})}\BibitemShut {NoStop}%
\bibitem [{\citenamefont {Kitney}\ and\ \citenamefont {Rompelman}(1980)}]{kitney1980study}%
  \BibitemOpen
  \bibfield  {author} {\bibinfo {author} {\bibfnamefont {R.}~\bibnamefont {Kitney}}\ and\ \bibinfo {author} {\bibfnamefont {O.}~\bibnamefont {Rompelman}},\ }\bibfield  {title} {\enquote {\bibinfo {title} {The study of heart-rate variability},}\ }\href@noop {} {\bibfield  {journal} {\bibinfo  {journal} {(No Title)}\ } (\bibinfo {year} {1980})}\BibitemShut {NoStop}%
\bibitem [{\citenamefont {Saul}\ \emph {et~al.}(1988)\citenamefont {Saul}, \citenamefont {Arai}, \citenamefont {Berger}, \citenamefont {Lilly}, \citenamefont {Colucci},\ and\ \citenamefont {Cohen}}]{saul1988assessment}%
  \BibitemOpen
  \bibfield  {author} {\bibinfo {author} {\bibfnamefont {J.~P.}\ \bibnamefont {Saul}}, \bibinfo {author} {\bibfnamefont {Y.}~\bibnamefont {Arai}}, \bibinfo {author} {\bibfnamefont {R.~D.}\ \bibnamefont {Berger}}, \bibinfo {author} {\bibfnamefont {L.~S.}\ \bibnamefont {Lilly}}, \bibinfo {author} {\bibfnamefont {W.~S.}\ \bibnamefont {Colucci}}, \ and\ \bibinfo {author} {\bibfnamefont {R.~J.}\ \bibnamefont {Cohen}},\ }\bibfield  {title} {\enquote {\bibinfo {title} {Assessment of autonomic regulation in chronic congestive heart failure by heart rate spectral analysis},}\ }\href@noop {} {\bibfield  {journal} {\bibinfo  {journal} {The American journal of cardiology}\ }\textbf {\bibinfo {volume} {61}},\ \bibinfo {pages} {1292--1299} (\bibinfo {year} {1988})}\BibitemShut {NoStop}%
\bibitem [{\citenamefont {Berger}\ \emph {et~al.}(1988)\citenamefont {Berger}, \citenamefont {Saul}, \citenamefont {Albrecht}, \citenamefont {Stein},\ and\ \citenamefont {Cohen}}]{berger1988respiratory}%
  \BibitemOpen
  \bibfield  {author} {\bibinfo {author} {\bibfnamefont {R.~D.}\ \bibnamefont {Berger}}, \bibinfo {author} {\bibfnamefont {J.~P.}\ \bibnamefont {Saul}}, \bibinfo {author} {\bibfnamefont {P.}~\bibnamefont {Albrecht}}, \bibinfo {author} {\bibfnamefont {S.~P.}\ \bibnamefont {Stein}}, \ and\ \bibinfo {author} {\bibfnamefont {R.~J.}\ \bibnamefont {Cohen}},\ }\bibfield  {title} {\enquote {\bibinfo {title} {Respiratory effects on arterial pressure: a novel signal analysis approach},}\ }in\ \href@noop {} {\emph {\bibinfo {booktitle} {Proceedings of the Annual International Conference of the IEEE Engineering in Medicine and Biology Society}}}\ (\bibinfo {organization} {IEEE},\ \bibinfo {year} {1988})\ pp.\ \bibinfo {pages} {533--534}\BibitemShut {NoStop}%
\bibitem [{\citenamefont {Saul}(1996)}]{saul1996transfer}%
  \BibitemOpen
  \bibfield  {author} {\bibinfo {author} {\bibfnamefont {J.~P.}\ \bibnamefont {Saul}},\ }\bibfield  {title} {\enquote {\bibinfo {title} {Transfer function analysis of cardiorespiratory variability to assess autonomic regulation},}\ }\href@noop {} {\bibfield  {journal} {\bibinfo  {journal} {Clinical Science}\ }\textbf {\bibinfo {volume} {91}},\ \bibinfo {pages} {101--101} (\bibinfo {year} {1996})}\BibitemShut {NoStop}%
\bibitem [{\citenamefont {Triedman}\ and\ \citenamefont {Saul}(1994)}]{triedman1994blood}%
  \BibitemOpen
  \bibfield  {author} {\bibinfo {author} {\bibfnamefont {J.~K.}\ \bibnamefont {Triedman}}\ and\ \bibinfo {author} {\bibfnamefont {J.~P.}\ \bibnamefont {Saul}},\ }\bibfield  {title} {\enquote {\bibinfo {title} {Blood pressure modulation by central venous pressure and respiration. buffering effects of the heart rate reflexes.}}\ }\href@noop {} {\bibfield  {journal} {\bibinfo  {journal} {Circulation}\ }\textbf {\bibinfo {volume} {89}},\ \bibinfo {pages} {169--179} (\bibinfo {year} {1994})}\BibitemShut {NoStop}%
\bibitem [{\citenamefont {De~Boer}, \citenamefont {Karemaker},\ and\ \citenamefont {Strackee}(1985{\natexlab{b}})}]{de1985relationshipsI}%
  \BibitemOpen
  \bibfield  {author} {\bibinfo {author} {\bibfnamefont {R.}~\bibnamefont {De~Boer}}, \bibinfo {author} {\bibfnamefont {J.}~\bibnamefont {Karemaker}}, \ and\ \bibinfo {author} {\bibfnamefont {J.}~\bibnamefont {Strackee}},\ }\bibfield  {title} {\enquote {\bibinfo {title} {Relationships between short-term blood-pressure fluctuations and heart-rate variability in resting subjects i: a spectral analysis approach},}\ }\href@noop {} {\bibfield  {journal} {\bibinfo  {journal} {Medical and biological engineering and computing}\ }\textbf {\bibinfo {volume} {23}},\ \bibinfo {pages} {352--358} (\bibinfo {year} {1985}{\natexlab{b}})}\BibitemShut {NoStop}%
\bibitem [{\citenamefont {DeBoer}, \citenamefont {Karemaker},\ and\ \citenamefont {Strackee}(1987)}]{deboer1987hemodynamic}%
  \BibitemOpen
  \bibfield  {author} {\bibinfo {author} {\bibfnamefont {R.~W.}\ \bibnamefont {DeBoer}}, \bibinfo {author} {\bibfnamefont {J.~M.}\ \bibnamefont {Karemaker}}, \ and\ \bibinfo {author} {\bibfnamefont {J.}~\bibnamefont {Strackee}},\ }\bibfield  {title} {\enquote {\bibinfo {title} {Hemodynamic fluctuations and baroreflex sensitivity in humans: a beat-to-beat model},}\ }\href@noop {} {\bibfield  {journal} {\bibinfo  {journal} {American Journal of Physiology-Heart and Circulatory Physiology}\ }\textbf {\bibinfo {volume} {253}},\ \bibinfo {pages} {H680--H689} (\bibinfo {year} {1987})}\BibitemShut {NoStop}%
\bibitem [{\citenamefont {Miyakawa}(1984)}]{miyakawa1984mechanism}%
  \BibitemOpen
  \bibfield  {author} {\bibinfo {author} {\bibfnamefont {K.}~\bibnamefont {Miyakawa}},\ }\bibfield  {title} {\enquote {\bibinfo {title} {Mechanism of blood pressure waves of the third order},}\ }\href@noop {} {\bibfield  {journal} {\bibinfo  {journal} {Mechanisms of blood pressure waves}\ ,\ \bibinfo {pages} {85--117}} (\bibinfo {year} {1984})}\BibitemShut {NoStop}%
\bibitem [{\citenamefont {Preiss}, \citenamefont {Iscoe},\ and\ \citenamefont {Polosa}(1975)}]{preiss1975analysis}%
  \BibitemOpen
  \bibfield  {author} {\bibinfo {author} {\bibfnamefont {G.}~\bibnamefont {Preiss}}, \bibinfo {author} {\bibfnamefont {S.}~\bibnamefont {Iscoe}}, \ and\ \bibinfo {author} {\bibfnamefont {C.}~\bibnamefont {Polosa}},\ }\bibfield  {title} {\enquote {\bibinfo {title} {Analysis of a periodic breathing pattern associated with mayer waves},}\ }\href@noop {} {\bibfield  {journal} {\bibinfo  {journal} {American Journal of Physiology-Legacy Content}\ }\textbf {\bibinfo {volume} {228}},\ \bibinfo {pages} {768--774} (\bibinfo {year} {1975})}\BibitemShut {NoStop}%
\bibitem [{\citenamefont {Saul}\ \emph {et~al.}(1991)\citenamefont {Saul}, \citenamefont {Berger}, \citenamefont {Albrecht}, \citenamefont {Stein}, \citenamefont {Chen},\ and\ \citenamefont {Cohen}}]{saul1991transfer}%
  \BibitemOpen
  \bibfield  {author} {\bibinfo {author} {\bibfnamefont {J.~P.}\ \bibnamefont {Saul}}, \bibinfo {author} {\bibfnamefont {R.~D.}\ \bibnamefont {Berger}}, \bibinfo {author} {\bibfnamefont {P.}~\bibnamefont {Albrecht}}, \bibinfo {author} {\bibfnamefont {S.}~\bibnamefont {Stein}}, \bibinfo {author} {\bibfnamefont {M.~H.}\ \bibnamefont {Chen}}, \ and\ \bibinfo {author} {\bibfnamefont {R.}~\bibnamefont {Cohen}},\ }\bibfield  {title} {\enquote {\bibinfo {title} {Transfer function analysis of the circulation: unique insights into cardiovascular regulation},}\ }\href@noop {} {\bibfield  {journal} {\bibinfo  {journal} {American Journal of Physiology-Heart and Circulatory Physiology}\ }\textbf {\bibinfo {volume} {261}},\ \bibinfo {pages} {H1231--H1245} (\bibinfo {year} {1991})}\BibitemShut {NoStop}%
\end{thebibliography}%

\pagebreak
\clearpage


\onecolumngrid
\textbf{\Large{Supplementary Materials for: Heart Rate Variability Series is the Output of a non-Chaotic System driven by Dynamical Noise}}

\vspace{0.5cm}
M. Bianco$^1$, A. Scarciglia$^1$, C. Bonanno$^2$, and G. Valenza$^1$

\vspace{0.5cm}
$^1)$ Dept. of Information Engineering and Research Centre E Piaggio, University of Pisa, Pisa, Italy 

$^2)$ Dept. of Mathematics, University of Pisa, Pisa, Italy




\setcounter{table}{0}
\renewcommand{\thetable}{S\arabic{table}}

\begin{table*}[b]
\caption{\label{tab:LogMap}Logistic map: in the first line of each cell the median ($\pm$ mad) of the values K computed for 100 series, in the second line the maximum value and the minimum value of K; if the interval [min(K) max(K)] crosses 0.9, the percentage of series with K less than 0.9 is reported in brackets.}
\begin{ruledtabular}
\begin{tabular}{|p{1.5cm}p{3.2cm}p{3.2cm}|p{1.5cm}p{3.2cm}p{3.2cm}|}
  \multicolumn{3}{|c|}{Logistic map (regular regime, $\mu=3.5$)}& \multicolumn{3}{c|}{Logistic map (chaotic regime, $\mu=3.97$)}\\ \hline
 sd noise & \ \ \ \ \ \ Dynamical noise& \ \ \ \ \ \ \ \  output noise & sd noise & \ \ \ \ \ \ Dynamical noise& \ \ \ \ \ \ \ \ output noise  \\ \hline
 
  0&   \multicolumn{2}{c|}{\makecell{$0.000 \pm$ 0.000\\ 0.000-(-0.002)}} &0 & \multicolumn{2}{c|}{\makecell{ 0.998 $\pm 0.000$ \\ 0.999-0.997}}\\ \hline
   
   2\%&  
  \makecell{$0.526 \pm$ 0.192\\ 0.799-0.154} & \makecell{$0.027 \pm 0.000$\\ 0.032-0.023} &  2\% & \makecell{$0.998 \pm$ 0\\0.999-0.997 } & \makecell{$0.998 \pm 0.000$\\0.998-0.995 } \\ \hline
  
   5\%  & \makecell{$0.674 \pm 0.110$\\0.990-0.461 (88\% <0.9) } & \makecell{$0.107 \pm 0.006$\\0.127-0.091} & 5\% & \makecell{$0.990 \pm 0.000$\\ 0.993-0.987 } & \makecell{$0.994 \pm 0.00a$\\ 0.996-0.975 } \\ \hline

  10\% & \makecell{$0.805 \pm 0.116 $\\ 0.988-0.509 (73\%<0.9)} & \makecell{$0.152 \pm 0.006$\\0.177-0.138 } &   10\% & \makecell{$0.967 \pm 0.003$\\0.975-0.961} & \makecell{$0.990 \pm 0.010$\\ 0.994-0.945} \\ \hline
 
  15\% & \makecell{$0.831 \pm 0.065$\\0.975-0.732 (68\%<0.9)} & \makecell{$0.267 \pm 0.009$\\0.303-0.237} & 15\% & \makecell{$0.953 \pm 0.003$\\0.962-0.945} & \makecell{$0.986 \pm 0.011$\\ 0.992-0.935} \\ \hline
 
 20\%  & \makecell{$0.861 \pm 0.039$\\ 0.957-0.806 (70\%<0.9)} & \makecell{$0.276 \pm 0.010$\\ 0.311-0.241} & 20\% & \makecell{$ 0.956\pm 0.003$\\ 0.962-0.943} & \makecell{$0.987 \pm 0.009$\\ 0.992-0.953} 
\end{tabular}
\end{ruledtabular}
\end{table*}
\vspace{-5cm}
\begin{table}[b]
\caption{\label{tab:PomMann}Pomeau-Manneville map:in the first line of each cell the median ($\pm$ mad) of the values K computed for 100 series, in the second line the maximum value and the minimum value of K.}
\begin{ruledtabular}
\begin{tabular}{|ccc|}
 \multicolumn{3}{|c|}{Pomeau-Mannevile map ($\gamma=0.2$)}\\ \hline
 sd noise &Dynamical noise& output noise\\ \hline
 0 & \multicolumn{2}{c|}{\makecell{ $0.998 \pm 0.000$ \\ 0.999-0.997}}\\ \hline
 
 2\%  & \makecell{$0.998 \pm 0.000$ \\ 0.998-0.996} & \makecell{$0.998 \pm 0.000$\\ 0.998-0.997} \\ \hline
 
 5\%  & \makecell{$0.989 \pm 0.002$\\ 0.993-0.985} & \makecell{$0.993 \pm 0.002$\\0.996-0.989} \\ \hline
 
 10\% & \makecell{$0.967 \pm 0.005 $\\ 0.979-0.958} & \makecell{$0.985 \pm 0.004$\\0.994-0.977 } \\ \hline
 
 15\%  & \makecell{$0.958 \pm 0.005$\\ 0.969-0.943} & \makecell{$0.984 \pm 0.004$\\ 0.991-0.975 } \\ \hline
 
 20\% & \makecell{$0.961 \pm 0.003$\\ 0.968-0.947} & \makecell{$0.985 \pm 0.002$\\ 0.991-0.980}\\
\end{tabular}
\end{ruledtabular}
\end{table}

\end{document}